\documentclass[10pt,twocolumn,letterpaper,pdftex,pdfa,pagebackref,breaklinks,colorlinks,allcolors=cvprblue]{article}

\usepackage[pagenumbers]{cvpr} 

\usepackage[a-2b]{pdfx}

\usepackage{adjustbox}
\usepackage{tikz}
\usetikzlibrary{calc,spy}

\usepackage{ifthen}
\usepackage{microtype}
\usepackage[moderate,lists=normal,mathspacing=normal,mathdisplays=normal]{savetrees}
\usepackage{multirow}

\usepackage{float}

\newcommand{\isdraft}{false}

\newboolean{revisedText}
\setboolean{revisedText}{true} 

\newcommand{\proposedChange}[2]{%
    \ifthenelse{\boolean{revisedText}}{\textcolor{black}{#1}}{#2}%
}

\expandafter\def\expandafter\normalsize\expandafter{%
    \normalsize%
    \setlength\abovedisplayskip{0pt plus 2pt}
    \setlength\belowdisplayskip{0pt plus 2pt}
    \setlength\abovedisplayshortskip{0pt plus 2pt}
    \setlength\belowdisplayshortskip{0pt plus 2pt}
}

\newcommand{\inlineheading}[1]{\vspace{6pt}\noindent\textbf{{#1}.}\hspace{0.5em}}

\newcommand{\phiQ}{Q_\phi}
\newcommand{\phiQSet}{\mathbb{\phiQ}}
\newcommand{\torf}{TöRF\xspace}
\newcommand{\ftorf}{F-TöRF\xspace}

\newcommand{\SeqPillow}{\emph{Pillow}\xspace}
\newcommand{\SeqBaseball}{\emph{Baseball}\xspace}
\newcommand{\SeqFan}{\emph{Fan}\xspace}

\newcommand{\SeqPhoneBooth}{\emph{PhoneBooth}\xspace}
\newcommand{\SeqCupboard}{\emph{Cupboard}\xspace}
\newcommand{\SeqPhotocopier}{\emph{Photocopier}\xspace}
\newcommand{\SeqStudyBook}{\emph{StudyBook}\xspace}
\newcommand{\SeqDeskBox}{\emph{DeskBox}\xspace}
\newcommand{\SeqJumpingJacks}{\emph{JumpingJacks}\xspace}
\newcommand{\SeqTarget}{\emph{Target}\xspace}
\newcommand{\SeqSlidingCube}{\emph{Sliding Cube}\xspace}
\newcommand{\SeqArcingCube}{\emph{Arcing Cube}\xspace}
\newcommand{\SeqThreeCubes}{\emph{3 Cubes Speed Test}\xspace}
\newcommand{\SeqThreeCubesShort}{\emph{3 Cubes ST}\xspace}
\newcommand{\SeqThreeChairs}{\emph{3 Chairs Speed Test}\xspace}
\newcommand{\SeqThreeChairsShort}{\emph{3 Chairs ST}\xspace}
\newcommand{\SeqOccludedCube}{\emph{Occluded Cube}\xspace}
\newcommand{\SeqAxial}{\emph{Axial Speed Test}\xspace}
\newcommand{\SeqAxialShort}{\emph{Axial ST}\xspace}
\newcommand{\SeqZMotion}{\emph{Orthogonal Speed Test}\xspace}
\newcommand{\SeqZMotionShort}{\emph{Ortho. ST}\xspace}

\usepackage{array}
\usepackage{colortbl}
\usepackage{makecell}
\usepackage{lipsum}

\definecolor{cvprblue}{rgb}{0.21,0.49,0.74}

\def\paperID{7710}
\def\confName{CVPR}
\def\confYear{2025}

\title{Time of the Flight of the Gaussians:\\Optimizing Depth Indirectly in Dynamic Radiance Fields}

\author{%
Runfeng Li\quad%
Mikhail Okunev\quad%
Zixuan Guo\quad%
Anh Ha Duong\quad%
\\[0.15em]%
Christian Richardt\textsuperscript{$\infty$}\quad%
Matthew O'Toole\textsuperscript{*}\quad%
James Tompkin%
\\[0.5em]%
Brown University\quad%
\textsuperscript{$\infty$} Meta Reality Labs\quad%
\textsuperscript{*} Carnegie Mellon University%
}

\begin{document}

\maketitle

\begin{abstract}
We present a method to reconstruct dynamic scenes from monocular continuous-wave time-of-flight (C-ToF) cameras using raw sensor samples that achieves similar or better accuracy than neural volumetric approaches and is 100$\times$ faster. 
Quickly achieving high-fidelity dynamic 3D reconstruction from a single viewpoint is a significant challenge in computer vision. 
In C-ToF radiance field reconstruction, the property of interest---depth---is not directly measured, causing an additional challenge.
This problem has a large and underappreciated impact upon the optimization when using a fast primitive-based scene representation like 3D Gaussian splatting, which is commonly used with multi-view data to produce satisfactory results and is brittle in its optimization otherwise.
We incorporate two heuristics into the optimization to improve the accuracy of scene geometry represented by Gaussians.
Experimental results show that our approach produces accurate reconstructions under constrained C-ToF sensing conditions, including for fast motions like swinging baseball bats.
\\
\href{https://visual.cs.brown.edu/gftorf}{https://visual.cs.brown.edu/gftorf}
\end{abstract}

\def\thefootnote{}\footnotetext{All experiments were performed by university authors.}\def\thefootnote{\arabic{footnote}}
  
\vspace{-12pt}
\section{Introduction}

Active illumination sensing, like continuous-wave time-of-flight (C-ToF), can help reconstruct dynamic scenes with a single camera thanks to their depth estimates. C-ToF cameras derive depth with simple reconstruction models that assume that all surfaces are opaque and Lambertian, which is fast but can lead to depth errors. Recent optimization-based approaches \cite{attal2021torf, okunev2024flowed} attempt to resolve the scene with more sophisticated physics that model the emitted light and its reflection from an underlying transmissive 4D volume. While too slow to optimize for practical use, so-called neural ToF radiance fields \cite{attal2021torf} are still promising because, in principle, they are better at modeling superposition effects from multi-path light transport.

\begin{figure}[t]
\vspace{-16pt}\centering
\includegraphics[width=0.95\columnwidth,clip,trim={0 370 480 0},page=1]{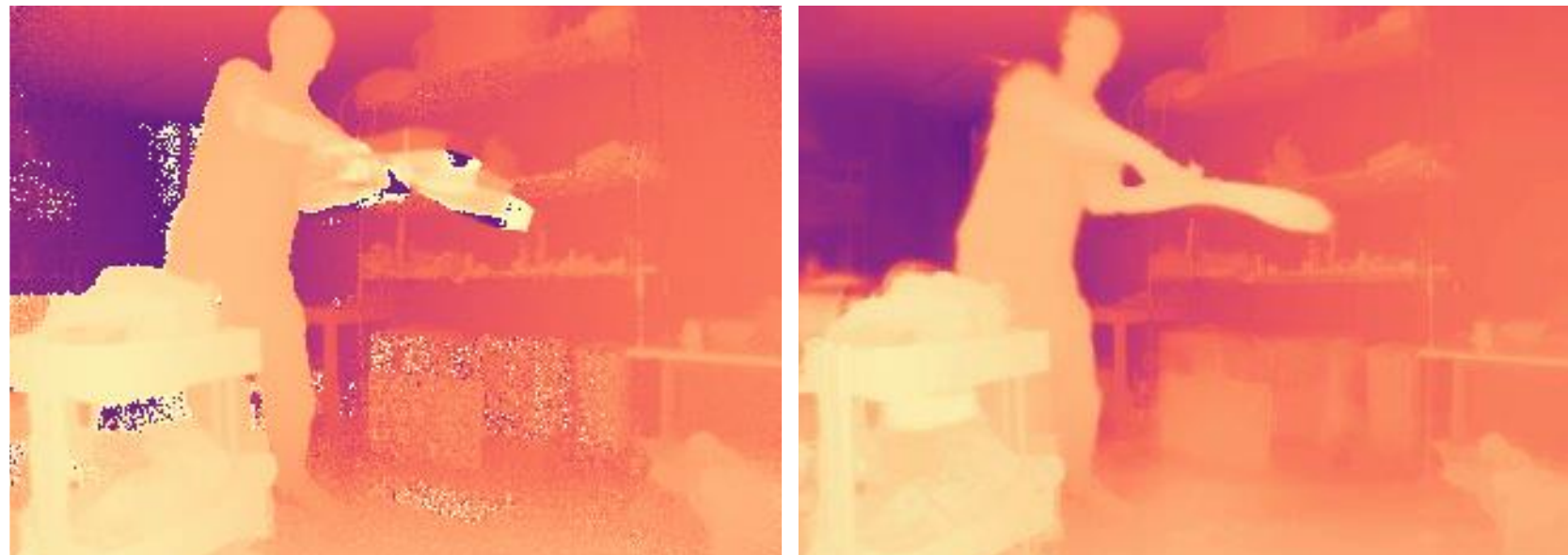}
\vspace{2pt}
\small{(a) C-ToF depth} \hspace{1.9cm} \small{(b) Our depth}
\includegraphics[width=0.95\columnwidth,clip,trim={0 235 480 192},page=1]{images/teaser.pdf}
\small{(c) 3D motion trajectories}
\vspace{-0.2cm}
\caption{\label{fig:teaser}
    Measuring the depth of a fast-moving object is challenging for C-ToF cameras.
    By modeling raw C-ToF frames, our method can reconstruct the geometry and motion of a fast swinging baseball bat. It is 100$\times$ faster to optimize and render and achieves similar or better accuracy than prior neural volumetric approaches.
}\vspace{-6pt}
\end{figure}

Our work considers two inter-related problems in this setting: how to make these methods faster, and how to make their optimizations more stable once we use faster---and brittler \cite{liang2024monocular}---reconstruction methods. By the end of this paper, we will have accuracy comparable to or better than existing methods while increasing speed by 100$\times$, so making single-camera 4D dynamic scene reconstruction more practical (\cref{fig:teaser}).

\begin{figure}[t]
    \centering
    \includegraphics[width=0.85\linewidth]{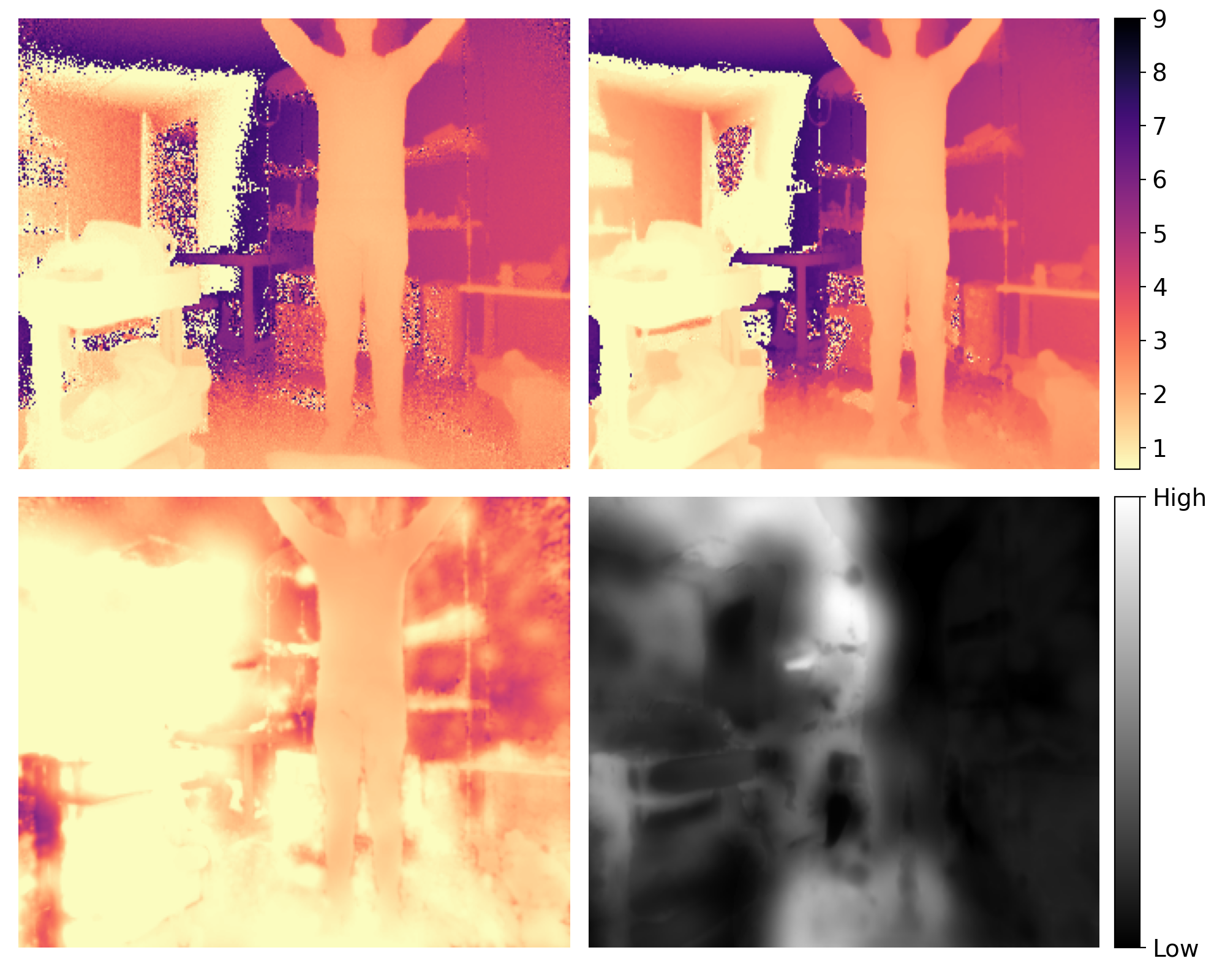}
    \vspace{-0.25cm}
    \caption{\textbf{Fitting C-ToF images $\neq$ fitting depth.} \emph{Top left:} Camera-derived depth from C-ToF. \emph{Top right:} Rendering a GS scene reconstruction into C-ToF raw image samples, then deriving depth. As this is similar to the camera-derived depth to the left, the reconstruction objective was met. 
    \emph{Bottom left:} Rendered mean scene depth from Gaussians, which is highly inaccurate.
    \emph{Bottom right:} Depth distortion error~\cite{huang20242d}, which measures the Gaussian sparsity along each ray. Gaussians are not well localized.}
    \label{fig:jensen_visual_explain}
\end{figure}

First, we explain why such an optimization might be unstable. For a single camera, even with active illumination, the accurate reconstruction of depth is not sufficiently constrained by C-ToF sensor measurements under the transmissive image formation model.
As depth is only ever indirectly optimized, the scene can have highly inaccurate depth but still produce high-quality reconstructions of sensor measurements (\cref{fig:jensen_visual_explain}).
Thus, as the problem is under-con\-strained, its optimization is sensitive to its initialization and hyperparameters\proposedChange{}{, and minor tweaks can cause large differences in the output}.
Past works have avoided this pernicious problem: \torf \cite{attal2021torf} uses additional constraints to localize the depth by both moving the camera and by integrating RGB images. This reduces the problem, but is not suitable for \proposedChange{static}{fixed (static)} cameras. \ftorf \cite{okunev2024flowed} provides a more challenging dataset with a static camera and fast-moving objects. But, without the additional constraints, its depth estimates can be worse than the simple C-ToF derived depth even for opaque surfaces.
Some works add priors on depth, e.g., using learned single-image depth \cite{zhu2023fsgs,li2024dngaussian} or priors over scenes \cite{charatan2024pixelsplat,xu2024grm}. While pragmatic, these approaches are tangential to our goal of attempting to use what sensing we have to accurately measure a scene.

Second, we explain how to speed up the optimization. One way to speed up NeRFs is to use a fast Gaussian splatting (GS) based approach \cite{kerbl20233d,yang2024deformable,liang2023gaufre}.
When there are sufficient cameras to constrain the optimization, GS methods can generally act as drop-in replacements for NeRFs.
But, in our setting, we have only a single camera with a dynamic scene, and our desired property of depth is only indirectly optimized by the reconstruction of the sensor measurements. 
This makes GS methods brittle, and it is difficult to produce accurate depth results for dynamic scenes with C-ToF imaging.
While ToF NeRFs are slow, due to their MLP-based scene representation, they are more robust to unfavourable initializations and hyperparameters within this under-constrained setting. 

Thus, the goal is to close this gap by making fast GS-based methods more robust in our challenging setting. \proposedChange{We show that, with careful tuning of the optimization process, we can reliably achieve convergence in the GS model while maintaining the ability to represent complex geometry and motion.}{Existing approaches often add losses to implicitly enforce opaque surface assumptions, but these are counter to our goal of modeling more complex light transport. Instead, we contribute an analysis and corresponding insight into how to better optimize the indirect measurement of depth within ToF radiance field reconstruction, using two simple heuristics. Then, we instantiate this insight into a dynamic scene reconstruction method for continuous-wave ToF imaging from phasors or raw quads using 3D Gaussian splatting. This improves optimization and rendering time by 100$\times$ over previous NeRF-based methods with comparable or better accuracy.}
To summarize, our contributions include:

\begin{itemize}
    \item A dynamic C-ToF GS method with 100× efficiency improvement in both optimization and rendering.
    \item An investigation into C-ToF radiance field optimization and the empirical biases critical for accurate reconstruction of C-ToF radiance fields.
\end{itemize}

\vspace{-5pt}
\inlineheading{Assumptions}
We assume that the emitter is co-located with the camera, which can cause shadowing errors on close objects. We assume that motion can be modeled as a piecewise linear function between timesteps; higher-order motion models may improve results for rotational motions. We use optical flow estimates within our optimization, which may have errors that propagate to the final results.

\section{C-ToF Imaging Principles}
Following \citet{okunev2024flowed}, C-ToF cameras illuminate the scene with a continuously-modulating amplitude of light, usually as a sinusoid $\sin(2\pi f t)$.%
\footnote{We can more accurately model light intensity with a non-negative signal $\tfrac{1}{2}\sin(2\pi f t)+\tfrac{1}{2}$; for clarity, we choose the simpler model.}
As light returns to the camera, it is correlated with a reference sinusoid to produce an image with pixel intensities $A \sin(\psi+\phi) + B$.
Here, $A$ represents the amount of light received at each pixel---the amplitude---and $B$ represents the bias and depends upon the ambient illumination.
Phase $\psi$ captures the time that light is in flight.
There is also a programmable temporal shift of the reference signal, $\phi$.
$A$, $B$, and $\psi$ are three unknowns, and so we need at least three intensity measurements captured at different offsets $\phi$.
Many cameras use four offsets for robustness, where $\phi \in \{ 0, \frac{\pi}{2}, \pi , \frac{3\pi}{2}\}$, and the camera produces a quartet of raw frames $\phiQSet = \{A \sin(\psi + \phi) + B\}$.

Given frames $\phiQSet$, a typical C-ToF camera recovers the phase $\psi$ and computes the distance by multiplying time traveled with the speed of light $c$ \cite{Hansard2012}:%
\begin{align}
    d_{\text{ToF}} = \frac{\mathrm{c}}{4 \pi f}\psi \hspace{0.75em} \text{where} \hspace{0.75em} \psi = \arctan\left( \frac{ Q_0 - Q_\pi }{ Q_\frac{\pi}{2} - Q_\frac{3\pi}{2} } \right) \text{,} \label{eq:tof_depth}\\
    \begin{gathered}
    \text{because} \hspace{1em} \arctan\left( \frac{ Q_0 - Q_\pi }{ Q_\frac{\pi}{2} - Q_\frac{3\pi}{2} } \right) = \\ 
    \arctan\left( \frac{A \sin(\psi) + B - (-A \sin(\psi) + B)}{A \cos(\psi) + B - (-A \cos(\psi) + B)}\right) =\\
    = \arctan\left(\tan(\psi)\right) = \psi \text{.} \nonumber
    \end{gathered}
\end{align}
We can represent the C-ToF signal as a complex phasor $a \cdot W(d_\text{path}) \!=\! a \cdot \exp{(\mathrm{i} \frac{2\pi d_\text{path}f}{\mathrm{c}})} \!=\! (Q_0-Q_\pi) + \mathrm{i} (Q_\frac{\pi}{2} - Q_\frac{3\pi}{2})$, where $a \!=\! 2A$ and $d_\text{path}$ is the light path length ($d_\text{path} \!=\! 2d_{\text{ToF}}$)~\cite{gupta2015phasor}. Then, the phase $\psi \!=\! \angle W(d_\text{path})$ and the amplitude $A \!=\! \frac{1}{2} |a\cdot W(d_\text{path})| \!=\! \frac{1}{2} \sqrt{ (Q_0 - Q_\pi)^2 + (Q_\frac{\pi}{2} - Q_\frac{3\pi}{2})^2 }$. Furthermore, suppose that $s$ is the emitted signal intensity, $r$ is the reflectivity of a surface in a certain direction, and $1/d^2_{\text{ToF}}$ is the emitted light falloff, then $A \!=\! s\cdot r / d^2_{\text{ToF}}$.

C-ToF cameras have limitations. First, they can only measure depth unambiguously up to $d_\text{u}=\frac{\mathrm{c}}{2f}$ since all depths over this range will map back to $[0, d_\text{u}]$ due to sinusoidal periodicity. Real cameras usually have $d_\text{u}$ in the range of 5–-10 m.
Second, as we must capture four frames, achieving 30\,Hz depth output requires raw frame capture at 120\,Hz, and deriving depth in this way assumes that the quartet $\phiQSet$ was captured simultaneously. This means that any motion within the quartet will cause depth errors as the light arriving at a pixel will come from different world points. 
Third, this model assumes that light reflects back to the sensor from a single surface within a vacuum, when in reality it travels in complex paths.

\section{C-ToF GS Image Formation}

Next, we explain how we extend Gaussian splatting methods for dynamic scene C-ToF radiance field reconstruction. We assume that the reader is familiar with both \citet{kerbl20233d} and monocular 4D extensions for single-camera dynamic scenes, e.g., \citet{yang2024deformable} or \citet{liang2023gaufre}. 

\inlineheading{GS image formation model}
Briefly, a scene is reconstructed by optimizing a large set of anisotropic Gaussians $\mathcal{G}_k \!\in\! \mathcal{G}$.
Each is characterized by its center 3D position $\mathbf{x}_k \!\in\! \mathbb{R}^3$, covariance matrix $\mathbf{\Sigma}_k$ describing its 3D scale and rotation, opacity $o_k \!\in\! [0,1]$, and view-dependent color $\mathbf{c}_k \!\in\! [0, 1]^3$ parameterized by 16$\times$3 spherical harmonic coefficients.
Given a rasterizer to form a Gaussian as its 2D projection $\mathcal{G}_k^{\text{2D}}$ on the sensor, the color at pixel \( \mathbf{x} \) in an output image is a weighted sum of contributing Gaussians:
\begin{align}
\mathbf{c}(\mathbf{x}) = \mathbf{c}^{\text{bg}} T_{N} + \sum_{k=1}^{N} \mathbf{c}_k o_k \mathcal{G}^{\text{2D}}_k(\mathbf{x}) \ T_k \text{,}
\label{eq:gs}
\end{align}

\noindent where $T_k = \prod_{l=1}^{k-1} (1 - o_l \mathcal{G}^{\text{2D}}_l(\mathbf{x}))$ and the background signal intensity \( \mathbf{c}^{\text{bg}} \) is scaled by the final transmittance $T_N$.

\inlineheading{C-ToF GS phasor formation model}
Next, we adapt \cref{eq:gs} to model C-ToF signals as phasors, following the image formation model from \citet{attal2021torf}.
Phasor imaging~\cite{gupta2015phasor} assumes that all $\phiQSet$ are captured simultaneously.
The phasor at pixel \( \mathbf{x} \) is given by:
\begin{align} 
\mathbf{p}(\mathbf{x}) = \mathbf{p}^{\text{bg}} T_{N} + \sum_{k=1}^{N} \frac{2sr_k}{d_k^2} W(d_k) \, o_k \, \mathcal{G}^{\text{2D}}_k(\mathbf{x}) \, T_k^2 \text{.}
\label{eq:GSToF_Image_Formation}
\end{align} 
We note the differences from left to right.
For the infrared C-ToF signal, rather than representing the scene radiance \( \mathbf{c}_k \) towards the camera directly as in \cref{eq:gs}, we must describe the emitted and returned light. We model the returned light as the product of the source intensity \( s \) and view-dependent surface reflectivity \( r_k \in [0,1] \) modeled with 16 spherical harmonic coefficients as a Gaussian property. This light undergoes an inverse square falloff \( 1/d_k^2 \), where \( d_k \) is the distance from each Gaussian center to the camera’s optical center.
Then, \( W(d_k) = \exp(\mathrm{i} \psi_k) \) models the light path importance, where the phase shift between the emitted and reflected light is \( \psi_k = \frac{4\pi d_k f}{\mathrm{c}} \), where \( f \) is the C-ToF camera modulation frequency and \( c \) is the speed of light. 
Finally, transmission \( T_k = \prod_{l=1}^{k-1} (1 - o_l \mathcal{G}^\text{2D}_l(\mathbf{x})) \) is now squared to represent light traveling to and from the scene through the radiance field.

\inlineheading{C-ToF GS raw image formation model}
Rather than a single phasor assumed to be captured at a single time, \citet{okunev2024flowed} expand this model to consider that the set of raw images $\phiQSet$ are captured over time.
To begin, we replace the phasor \( W(d_k) \) in \cref{eq:GSToF_Image_Formation} with the modulated raw sample quads using $ \phi{(d_k)} : \{\sin{(\psi_k)}, \cos{(\psi_k)}, -\sin{(\psi_k)}, -\cos{(\psi_k)}\}$:
\begin{align} 
\mathbf{q}(\mathbf{x}) = \mathbf{q}^{\text{bg}} T_{N} + \sum_{k=1}^{N} \frac{s r_k}{d_k^2} \phi(d_k) \, o_k \, \mathcal{G}^\text{2D}_k(\mathbf{x}) \, T_k^2 \text{.}
\label{eq:GSRaw_Image_Formation}
\end{align}
Given a dynamic scene, this lets us render each raw quad $Q$ at an arbitrary time $t$. Then, given a real asynchronous $\phiQSet$, we must reconstruct the scene by finding correspondence across time between $Q$ such that we can resolve depth even though the scene is moving. \citet{okunev2024flowed} optimize for this correspondence while also solving for the scene's geometry.

\subsection{Where the Problems Lie}
\vspace{-5pt}
\inlineheading{C-ToF samples are not depth}
Unlike other problem settings that directly optimize the property of interest, e.g., color images for novel view synthesis or depth maps for depth estimation \cite{kerbl20233d, li2024dngaussian, chung2024depth}, our challenge is that the optimized property---C-ToF measurements, either as phasors (\cref{eq:GSToF_Image_Formation}) or as raw quads (\cref{eq:GSRaw_Image_Formation})---are not the same as the property of interest (depth).
Suppose that we compute the depth of the scene as the mean Gaussian depth, as given by:
\begin{align} 
d(\mathbf{x}) = \sum_{k=1}^{N} d_k \, o_k \, \mathcal{G}^\text{2D}_k(\mathbf{x}) \, T_k \text{.}
\label{eq:Depth_Image_Formation}
\end{align}
Without additional constraints, fitting \cref{eq:GSToF_Image_Formation} or \cref{eq:GSRaw_Image_Formation} volumetrically does not guarantee consistency with the depth rendered by \cref{eq:Depth_Image_Formation}. For example, when multiple density peaks exist along a ray but only a single surface exists, the rendered mean depth will diverge from the depth derived from the raw quads \emph{even if} the raw quads are accurately reconstructed (\cref{fig:jensen_visual_explain}). This discrepancy arises because the sine and cosine functions in \( \phi(d_k) \) oscillate between convex and concave regions, preventing equality in the finite form of Jensen's inequality. Only when all Gaussians are clustered at the same \( z \)-position along each ray does this depth gap reduce. 

\inlineheading{Previous approaches}
Past works have integrated additional multi-view constraints from moving cameras or extra RGB input~\cite{attal2021torf} to alleviate it. Of course, this works with a Gaussian renderer as well, but such an approach avoids the problem rather than finds a better solution to it. Further, experimentally, previous NeRF-based C-ToF radiance field methods \cite{attal2021torf, okunev2024flowed} exhibit this gap to a lesser extent, possibly due to the MLP’s inherent bias towards low-entropy (simple) solutions: a single high-density spike (a surface) has lower entropy along the ray than many locations with mid- or low-density values.

In contrast, GS has no such inherent bias; Gaussians can be positioned arbitrarily along the ray to fit the target, which harms depth accuracy. Supposing that we start from randomly initialized Gaussians, the flexibility in Gaussian reflectivities along each ray allows the optimizer to fit C-ToF data by forming high-entropy, multi-peak density distributions. Although this approach still causes a rendering of the scene to match the C-ToF data, the scene diverges from our goal of producing accurate mean depth maps.

\begin{figure*}[t]
    \includegraphics[trim=0 270 65 0, clip, width=\linewidth]{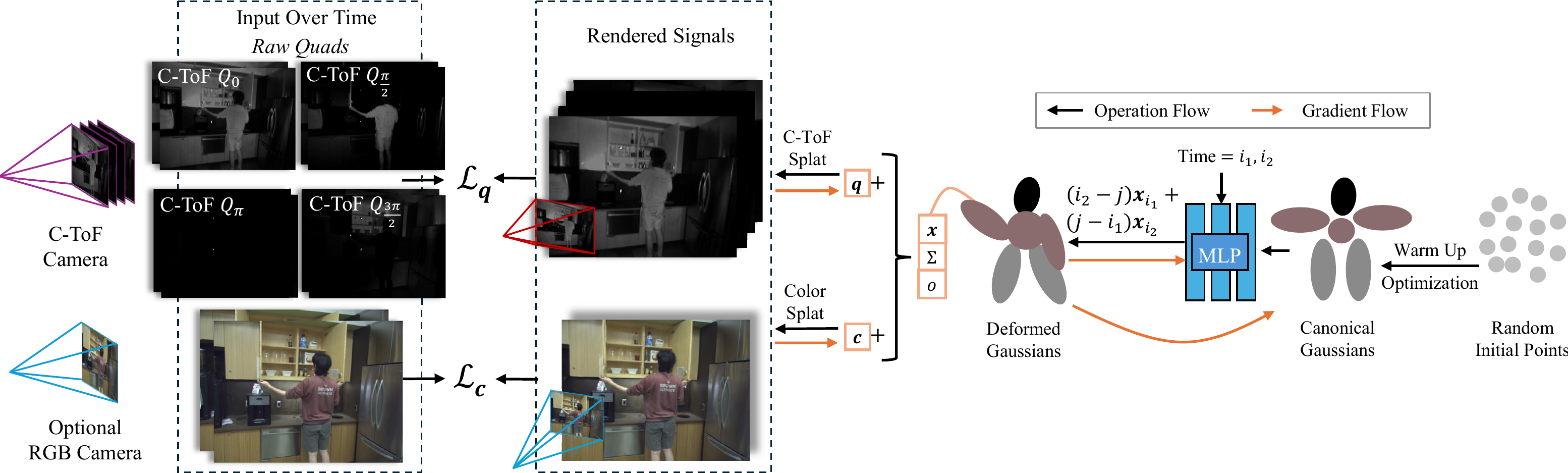}
    \vspace{-0.5cm}
    \caption{\label{fig:pipeline}
        \textbf{Pipeline.} \emph{Left:} We capture input raw quads (or phasors, not shown) from a continuous-wave time-of-flight camera with optional color camera. \emph{Right to Left:} From randomly initialized Gaussians, the warm-up stage estimates canonical scene geometry for a static scene.
        Then, given time $t$, the MLP predicts offsets ($\delta \mathbf{x}_k$) that reposition the canonical Gaussians.
        Then, we render the C-ToF and color images and compute losses.}
    \vspace{-0.1cm}
\end{figure*}

Current approaches in RGB scene reconstruction---even in more-constrained multi-view settings---use a depth distortion loss to help push Gaussians towards a surface~\cite{huang20242d}:
\begin{equation}
\text{DD}(\mathbf{x}) = \sum_{k=1}^N\sum_{l=1}^N \omega_k\omega_l\|d_k-d_l\|^2 \text{,}
\label{eq:ddloss}
\end{equation}
where $\omega_k \!=\! o_k \, \mathcal{G}^\text{2D}_k(\mathbf{x}) \, T_k$.
This implicitly introduces an opaque surface assumption, and often the scenes reconstructed by these methods do conform to this assumption.
However, this implicit assumption reduces our ability to model complex light transport. If we only wanted to reconstruct opaque surfaces, we would just derive depth from C-ToF in a closed form. The loss also tends to produce oversmoothed depth reconstructions.

\inlineheading{Heuristic 1: Occupancy bias}
To address this problem, we note that, within the space of reconstruction parameters and mechanisms when trying to reconstruct scenes (reflectivity, position, opacity, densification), `where things are' or `occupancy' is the primary variance, and so reflectivity changes should be rarer than occupancy changes (position, opacity, densification). Given the choice, we would rather move points around, make new density, or remove density than change the proportion of light that is returned.
To achieve this, we can decrease the learning rate 
within the optimization 
for scene reflectivity 
10$\times$, which increases the effect of position and opacity variation instead. While simple, this adjustment better encourages Gaussians to conform to the true scene structure from the outset \emph{without} forcing Gaussians to be close to each other as in \cref{eq:ddloss}. 

\inlineheading{Heuristic 2: Low-reflectivity bias}
The initial reflectivity plays a crucial role in the optimization result, particularly for low-reflectivity regions of the scene.
Assume that initial reflectivity values are high (e.g., 0.5). 
Then, with an occupancy bias, due to the inverse-square falloff term, we encourage Gaussians for low-reflectivity regions to be incorrectly placed far away from the camera.
Before the reflectivity of these far Gaussians can be corrected, and as Gaussians have spatial extent once splatted, other closer splat-overlapping Gaussians interfere to prevent the reflectivity correction. This leads to multiple opacity peaks. 

Now let us assume a lower initial reflectivity (0.01~to~0.1).
This enables Gaussians for low-reflectivity regions to be placed more correctly because the lower initialization provides a closer starting fit to the true scene.
With an occupancy bias, then Gaussians for high-reflectivity regions will be placed closer to the camera during initial optimization stages also due to the inverse-square falloff term.
As they occlude the scene, they are less likely to interfere with other Gaussians, and so have more iterations to optimize their reflectivities and positions to reach the true depth.

Empirically, combining occupancy and low-reflectivity biases is effective for both low and high reflectivity scene regions in our test sequences, whereas only using an occupancy bias (no low-reflectivity bias) harms low-reflectivity regions. 
While simple, our approach is more effective than \cref{eq:ddloss} without overly constraining the final result. Our supplemental document and video show evidence for this finding.

\section{Pipeline} 

As input, our system takes a video sequence of raw quartet C-ToF images and an optional color sequence (\cref{fig:pipeline}).
Our approach equivalently works with C-ToF phasors. 
We assume accurate camera poses, which align the scene scale with the C-ToF measurements. For our dynamic GS approach~\cite{yang2024deformable}, we use an MLP that takes time \( t \) and Gaussian 3D positions \( \mathbf{x}_k \) as input and outputs Gaussian 3D position offsets \( \delta \mathbf{x}_k \). Thus, Gaussians are implicitly deformed from a canonical space. 

Local linearity within each four raw frames is critical for fitting as the quartet is asynchronously captured. Let us consider an image in a quartet to be at an `integer' timestep $i$, and its matching-amplitude raw quad in four ticks time to be at $i+1$. Then we only deform Gaussians to integer time steps but ensure local linearity at intermediate fractional time steps \( j \) by linearly interpolating Gaussian positions from the two nearest integer timesteps \( i_1 \) and \( i_2 \): 
\begin{equation}
    \mathbf{x}_{j} = (i_2 - j)\mathbf{x}_{i_1} + (j - i_1)\mathbf{x}_{i_2} \text{.}
    \label{eq:linear_motion_intr}
\end{equation}
This is functionally equivalent to the phase-aware reprojection loss from \citet{okunev2024flowed}.

\inlineheading{Optical flow weak supervision}
To help correspond fast motions, we also weakly supervise the forward and backward 3D motion offsets of Gaussians against estimated optical flow from RAFT \cite{teed2020raft} with an L2 loss $\mathcal{L}_\text{f}$.
Optical flow is projected from the estimated 3D position offsets, where the forward flow is computed as (similarly for backward flow):
\begin{align}
    \mathbf{f}(\mathbf{x}) = \Pi_i\left(\mathrm{sg}(\mathbf{x}^d) + \sum_{k=1}^N \Delta \mathbf{x}_k o_k \mathcal{G}^{\text{2D}}_k(\mathbf{x}) T_k\right),
\end{align}
where \( \Delta \mathbf{x}_k = \text{MLP}(\mathbf{x}, i+1) - \text{MLP}(\mathbf{x}, i) \) is the Gaussian 3D motion, \( \mathbf{x}^d \) is the back-projected 3D point from the rendered mean depth, \( \mathrm{sg} \) stops gradient computation, and \( \Pi_i \) is the camera’s perspective projection at time \( i \).

\inlineheading{Total loss}
We combine three objectives:
\begin{align}
    \mathcal{L} = \alpha \mathcal{L}_\text{q} + \mathcal{L}_\text{c} + \beta \mathcal{L}_\text{f} \text{,}
\end{align}
where \( \mathcal{L}_\text{q} \) is the raw quad reconstruction loss and \( \mathcal{L}_\text{c} \) is the optional color reconstruction loss.
Both use \( \mathcal{L}_{2} \) and SSIM.

\inlineheading{Initialization}
We initialize Gaussians randomly within the camera frustum, bounded by near and far planes, and with a reflectivity of 0.1.
Unlike other dynamic GS models, our Gaussian positions are in real-world units, potentially with large values. To ease optimization, we rescale input coordinates to the MLP (only) using the unambiguous depth range, ensuring most values lie within [0, 1]. 
Further, in an initial warm-up stage for 2\,K iterations, we assume that the scene is static. This helps to place many of the Gaussians into useful positions. 

\vspace{0.1cm}
\inlineheading{Random background}
For sensitive low-reflectivity scene areas, a low background contribution $\mathbf{q}^{\text{bg}}$ can lead to errors: Gaussians should be placed on a surface to contribute, but the background is already mostly sufficient to reproduce the C-ToF signal. This creates only low-contributing Gaussians. To fix this, we randomly change the background signal (uniformly between –1 and 1) at each optimization iteration.

\vspace{0.1cm}
\inlineheading{MLP}
The deformation MLP consists of 8 layers with 256 neurons each, using positional encoding frequencies of \( L_\text{x} \!=\! 10 \) and \( L_\text{t} \!=\! 10 \). 
MLP biases are initialized to zero, and weights are initialized with Xavier normal initialization, except for the final linear layer, which maps the final 256 features to the delta position. 
This layer’s weights are initialized from a normal distribution with a standard deviation of $10^{-5}$ to ensure stable optimization by starting Gaussian deformations from small \( \delta \mathbf{x}_k \) values. 

\vspace{0.1cm}
\inlineheading{Hyperparameters}
In the loss, we set \( \alpha = 5 \) (\( 1 \) for synthetic scenes) and \( \beta = 0.0008 \).
We use the Adam optimizer with \( \beta = (0.9, 0.999) \) and \( \epsilon = 10^{-15} \).
We initialize reflectivity as \( 0.1 \) and use a reflectivity learning rate of $\gamma_\text{r}=0.00016$. 
The MLP learning rate starts at $\gamma_\text{MLP}=0.0008$ and decays exponentially to \( 1.6 \times 10^{-6} \) over 60\,K iterations.

\vspace{0.1cm}
\inlineheading{Implementation details}
Our code is based on \citet{kerbl20233d}, with custom CUDA implementations for forward and backward raw C-ToF image rasterization. Details on C-ToF Gaussian gradient calculations are provided in the supplemental material. Gaussians are rendered twice using a differentiable rasterizer: once for the C-ToF view to compute the quad loss, and once for the RGB view to compute the color loss.
\section{Experiments}

\begin{table*}[t]
    \caption{\textbf{\label{quantitative-synthetic-table} Depth error on the \ftorf synthetic dataset}. Each number is a depth MSE$\times 100$.
    $d$ is a standard volume-integrated depth from \cref{eq:Depth_Image_Formation}.
    $d_\text{ToF}$ is depth derived from the reconstructed C-ToF raw images using \cref{eq:tof_depth}. Bold marks best result for both $d$ and $d_\text{ToF}$.} 
    \vspace{-6pt}
    \centering
    \small
    \begin{tabular}{l | r | r | r | r | r | r | r}
        \toprule
         & {\SeqSlidingCube} & {\SeqOccludedCube} & {\SeqAxialShort} & {\SeqThreeCubesShort} & {\SeqThreeChairsShort} & {\SeqArcingCube} & {\SeqZMotionShort} \\
        \midrule
        C-ToF & {0.096} & {0.130} & {2.068} & {3.131} & {0.787} & {36.768} & {41.931} \\
        2D Flow & {0.018} & {0.088} & {0.662} & {0.916} & {\textbf{0.229}} & {1.678} & {19.805} \\
        \arrayrulecolor{black!30}\midrule
        \ftorf \(d_{\text{ToF}}\) & {0.023} & {0.114} & {0.251} & {0.501} & {0.324} & {\textbf{0.470}} & {27.488} \\
        Ours \(d_{\text{ToF}}\) & {\textbf{0.005}} & {\textbf{0.062}} & {\textbf{0.123}} & {\textbf{0.281}} & {0.639} & {1.060} & {\textbf{14.933}} \\
        \arrayrulecolor{black!100}\midrule
        \midrule
        \torf \(d\) & {0.349} & {0.388} & {1.143} & {2.363} & {0.956} & {6.278} & {22.855} \\
        \ftorf \(d\) & {0.440} & {0.647} & {0.938} & {1.390} & \bf{0.855} & {1.256} & {\textbf{7.527}} \\
        Ours \(d\) & {\textbf{{0.037}}} & {\textbf{{0.369}}} & {\textbf{{0.525}}} & {\textbf{{0.641}}} & {1.023} & {\textbf{{1.023}}} & {22.772} \\
        \bottomrule
    \end{tabular}
\end{table*}

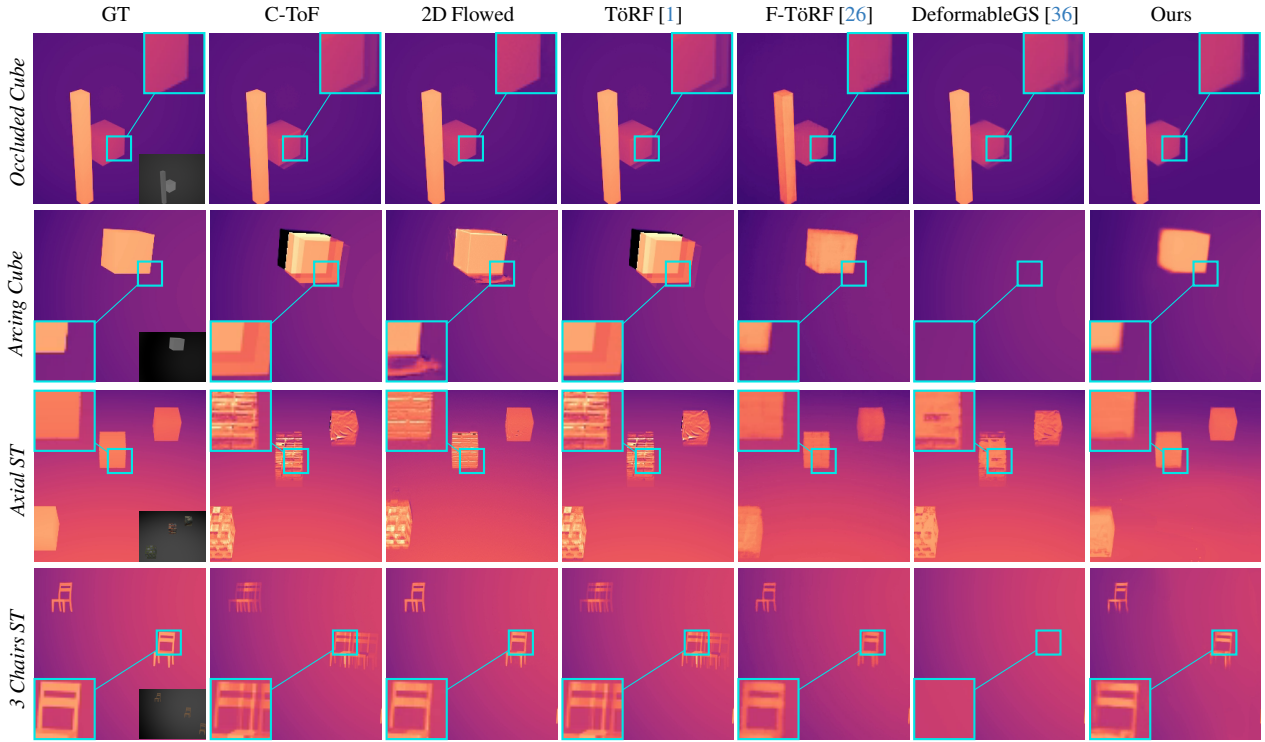
\begin{figure*}
    \centering
    \footnotesize

    \newcommand{\imgw}{0.13\linewidth}
    \newcommand{\imgwtiny}{0.050\linewidth} 
    
    \newcommand{\zoomxOcclusion}{0.0} 
    \newcommand{\zoomyOcclusion}{0.4} 

    \newcommand{\zoomxArcingCube}{-0.4} 
    \newcommand{\zoomyArcingCube}{-0.3} 

    \newcommand{\zoomxZMotionSpeedTest}{0.0} 
    \newcommand{\zoomyZMotionSpeedTest}{-0.2} 

    \newcommand{\zoomxSpeedTestChair}{-0.65} 
    \newcommand{\zoomySpeedTestChair}{-0.15} 

    \newcommand{\imgDepth}[2][0px 0px 0px 0px]{\includegraphics[draft=\isdraft,width=\imgw,trim=#1,clip,viewport=50 0 290 240]{#2}}
    \newcommand{\zoomDepth}[4]{
        \trimbox{0cm 0cm 0.15cm 0.0cm}{
        \begin{tikzpicture}[
    		image/.style={inner sep=0pt, outer sep=0pt},
    		collabel/.style={above=9pt, anchor=north, inner ysep=0pt, align=center}, 
    		rowlabel/.style={left=9pt, rotate=90, anchor=north, inner ysep=0pt, scale=0.8, align=center},
    		subcaption/.style={inner xsep=0.75mm, inner ysep=0.75mm, below right},
    		arrow/.style={-{Latex[length=2.5mm,width=4mm]}, line width=2mm},
    		spy using outlines={rectangle, size=0.8cm, magnification=2.5, connect spies, ultra thick, every spy on node/.append style={thick}},
    		style1/.style={cyan!90!black,thick},
    		style2/.style={orange!90!black},
    		style3/.style={blue!90!black},
    		style4/.style={green!90!black},
    		style5/.style={white},
    		style6/.style={black},
        ]
        
        \node [image] (#1) {\imgDepth[0px 0px 0px 0px]{#2}};
        \spy[style1] on ($(#1.center)-#3$) in node (crop-#1) [anchor=#4] at (#1.#4);
        
        \end{tikzpicture}
        }
    }

    \newcommand{\imgRGBTiny}[2][0px 0px 0px 0px]{\includegraphics[draft=\isdraft,width=\imgwtiny,trim=#1,clip]{#2}}
    \newcommand{\zoomDepthTinyRGB}[5]{
        \trimbox{0cm 0cm 0.15cm 0.0cm}{
        \begin{tikzpicture}[
    		image/.style={inner sep=0pt, outer sep=0pt},
                image2/.style={inner sep=0pt, outer sep=0pt},
    		collabel/.style={above=9pt, anchor=north, inner ysep=0pt, align=center},
    		rowlabel/.style={left=9pt, rotate=90, anchor=north, inner ysep=0pt, scale=0.8, align=center},
    		subcaption/.style={inner xsep=0.75mm, inner ysep=0.75mm, below right},
    		arrow/.style={-{Latex[length=2.5mm,width=4mm]}, line width=2mm},
    		spy using outlines={rectangle, size=0.8cm, magnification=2.5, connect spies, ultra thick, every spy on node/.append style={thick}},
    		style1/.style={cyan!90!black,thick},
    		style2/.style={orange!90!black},
    		style3/.style={blue!90!black},
    		style4/.style={green!90!black},
    		style5/.style={white},
    		style6/.style={black},
        ]
        
        \node [image] (#1) {\imgDepth[0px 0px 0px 0px]{#2}};
        \spy[style1] on ($(#1.center)-#3$) in node (crop-#1) [anchor=#4] at (#1.#4);

        \node [image2] at (-19.8pt,22.8pt) {\imgRGBTiny[0px 0px 0px 0px]{#5}};
        
        \end{tikzpicture}
        }
    }

    \newcommand{\zoomDepthTinyRGBbottom}[5]{
        \trimbox{0cm 0cm 0.15cm 0.0cm}{
        \begin{tikzpicture}[
    		image/.style={inner sep=0pt, outer sep=0pt},
                image2/.style={inner sep=0pt, outer sep=0pt},
    		collabel/.style={above=9pt, anchor=north, inner ysep=0pt, align=center},
    		rowlabel/.style={left=9pt, rotate=90, anchor=north, inner ysep=0pt, scale=0.8, align=center},
    		subcaption/.style={inner xsep=0.75mm, inner ysep=0.75mm, below right},
    		arrow/.style={-{Latex[length=2.5mm,width=4mm]}, line width=2mm},
    		spy using outlines={rectangle, size=0.8cm, magnification=2.5, connect spies, ultra thick, every spy on node/.append style={thick}},
    		style1/.style={cyan!90!black,thick},
    		style2/.style={orange!90!black},
    		style3/.style={blue!90!black},
    		style4/.style={green!90!black},
    		style5/.style={white},
    		style6/.style={black},
        ]
        
        \node [image] (#1) {\imgDepth[0px 0px 0px 0px]{#2}};
        \spy[style1] on ($(#1.center)-#3$) in node (crop-#1) [anchor=#4] at (#1.#4);

        \node [image2] at (20pt,-22.9pt) {\imgRGBTiny[0px 0px 0px 0px]{#5}};
        
        \end{tikzpicture}
        }
    }

    \setlength{\tabcolsep}{0.5pt}
    \begin{tabular}{cccccccc}
    &
    GT &
    C-ToF &
    2D Flowed &
    \torf \cite{attal2021torf} &
    \ftorf \cite{okunev2024flowed} &
    DeformableGS \cite{yang2024deformable} &
    Ours
    \\[0.2em]
    
    {\raisebox{0.4in}{\rotatebox[origin=c]{90}{\SeqOccludedCube}}} &
    \zoomDepthTinyRGBbottom{label0}{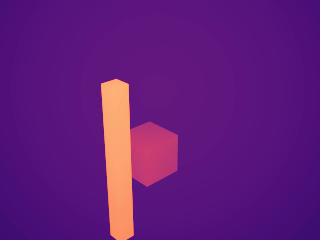}{(\zoomxOcclusion,\zoomyOcclusion)}{north east}
    {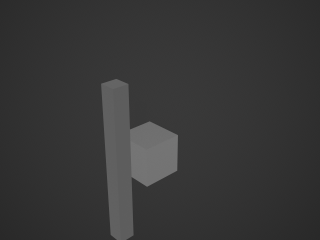}
    &
    \zoomDepth{label1}{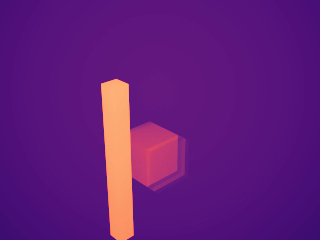}{(\zoomxOcclusion,\zoomyOcclusion)}{north east} &
    \zoomDepth{label2}{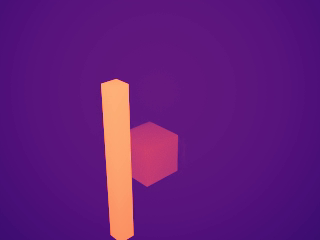}{(\zoomxOcclusion,\zoomyOcclusion)}{north east} &
    \zoomDepth{label3}{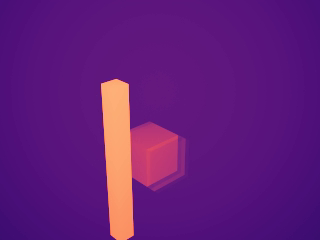}{(\zoomxOcclusion,\zoomyOcclusion)}{north east} &
    \zoomDepth{label4}{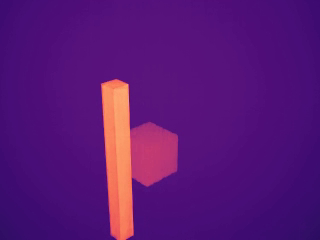}{(\zoomxOcclusion,\zoomyOcclusion)}{north east} &
    \zoomDepth{label5}{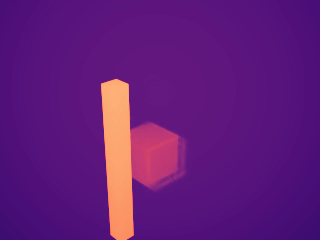}{(\zoomxOcclusion,\zoomyOcclusion)}{north east} &
    \zoomDepth{label5}{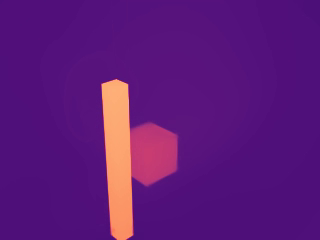}{(\zoomxOcclusion,\zoomyOcclusion)}{north east} \\
    
    {\raisebox{0.4in}{\rotatebox[origin=c]{90}{\SeqArcingCube}}} &
    \zoomDepthTinyRGBbottom{label0}{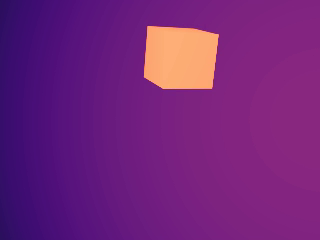}{(\zoomxArcingCube,\zoomyArcingCube)}{south west}
    {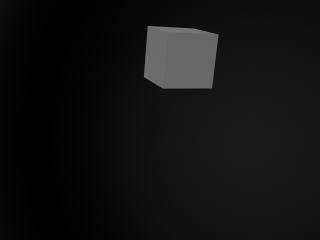}
    &
    \zoomDepth{label1}{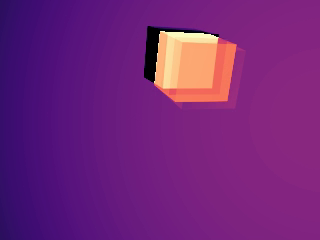}{(\zoomxArcingCube,\zoomyArcingCube)}{south west} &
    \zoomDepth{label2}{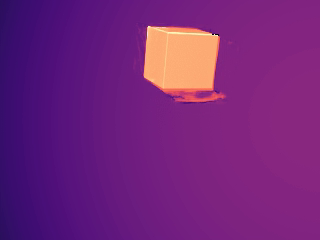}{(\zoomxArcingCube,\zoomyArcingCube)}{south west} &
    \zoomDepth{label3}{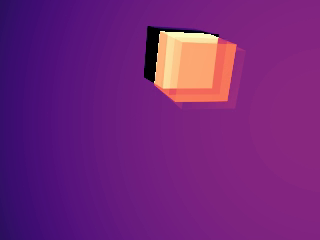}{(\zoomxArcingCube,\zoomyArcingCube)}{south west} &
    \zoomDepth{label4}{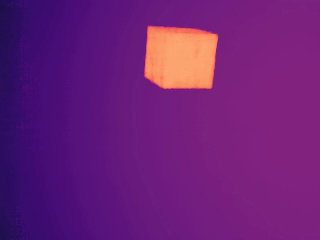}{(\zoomxArcingCube,\zoomyArcingCube)}{south west} &
    \zoomDepth{label5}{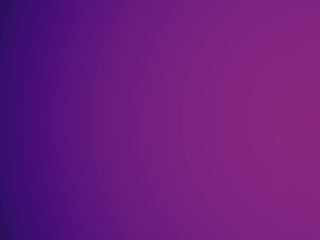}{(\zoomxArcingCube,\zoomyArcingCube)}{south west} &
    \zoomDepth{label5}{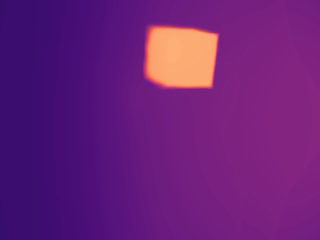}{(\zoomxArcingCube,\zoomyArcingCube)}{south west} \\

    {\raisebox{0.4in}{\rotatebox[origin=c]{90}{\SeqAxialShort}}} &
    \zoomDepthTinyRGBbottom{label1}{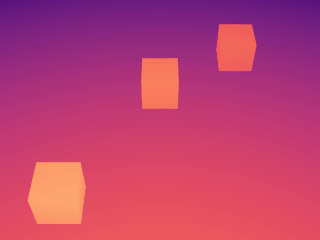}{(\zoomxZMotionSpeedTest,\zoomyZMotionSpeedTest)}{north west} 
    {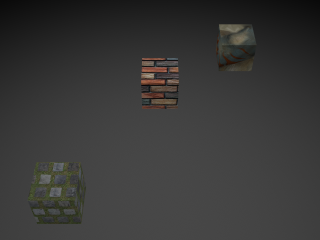}
    &
    \zoomDepth{label1}{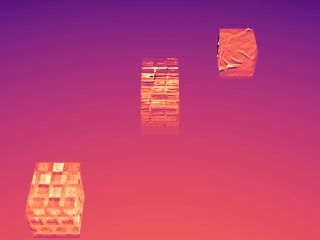}{(\zoomxZMotionSpeedTest,\zoomyZMotionSpeedTest)}{north west} &
    \zoomDepth{label2}{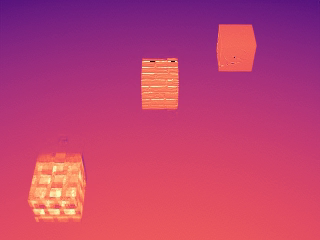}{(\zoomxZMotionSpeedTest,\zoomyZMotionSpeedTest)}{north west} &
    \zoomDepth{label3}{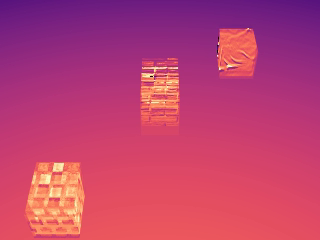}{(\zoomxZMotionSpeedTest,\zoomyZMotionSpeedTest)}{north west} &
    \zoomDepth{label4}{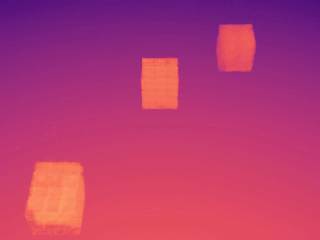}{(\zoomxZMotionSpeedTest,\zoomyZMotionSpeedTest)}{north west} &
    \zoomDepth{label5}{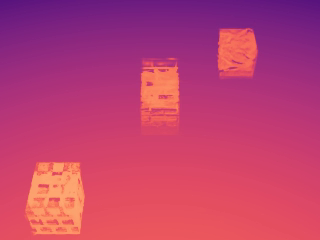}{(\zoomxZMotionSpeedTest,\zoomyZMotionSpeedTest)}{north west} &
    \zoomDepth{label5}{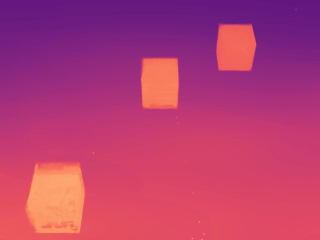}{(\zoomxZMotionSpeedTest,\zoomyZMotionSpeedTest)}{north west} \\

    {\raisebox{0.4in}{\rotatebox[origin=c]{90}{\SeqThreeChairsShort}}} &
    \zoomDepthTinyRGBbottom{label0}{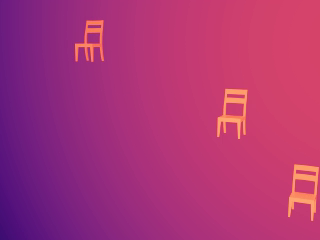}{(\zoomxSpeedTestChair,\zoomySpeedTestChair)}{south west} 
    {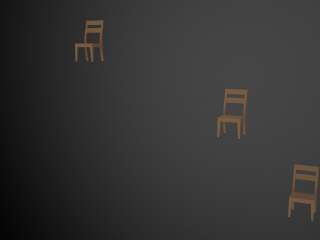}
    &
    \zoomDepth{label1}{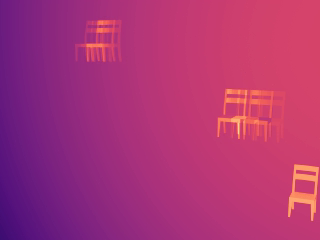}{(\zoomxSpeedTestChair,\zoomySpeedTestChair)}{south west} &
    \zoomDepth{label2}{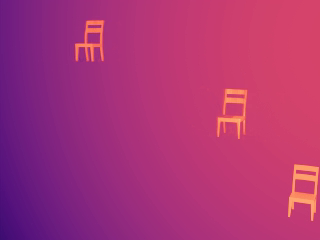}{(\zoomxSpeedTestChair,\zoomySpeedTestChair)}{south west} &
    \zoomDepth{label3}{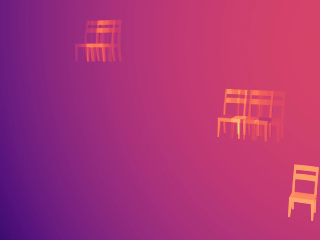}{(\zoomxSpeedTestChair,\zoomySpeedTestChair)}{south west} &
    \zoomDepth{label4}{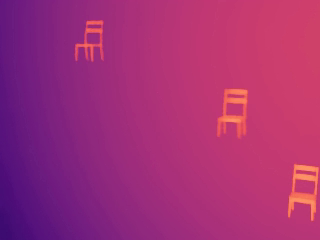}{(\zoomxSpeedTestChair,\zoomySpeedTestChair)}{south west} &
    \zoomDepth{label5}{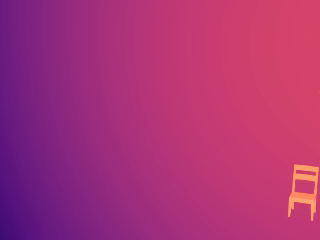}{(\zoomxSpeedTestChair,\zoomySpeedTestChair)}{south west} &
    \zoomDepth{label5}{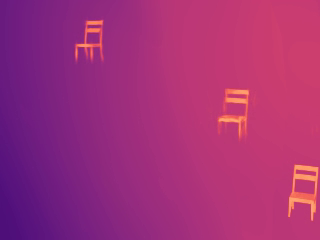}{(\zoomxSpeedTestChair,\zoomySpeedTestChair)}{south west} \\
    
    \end{tabular}
    \vspace{-3mm}
    \caption{\textbf{Our approach is competitive or better in terms of accuracy against the state of the art on synthetic scenes, while being two orders of magnitude faster.}
    We use the synthetic dataset from \ftorf to demonstrate this.
    All images show rendered volumetric depth $d$.
    State-of-the-art NeRF model \ftorf produces similar results to ours while being significantly slower.
    DeformableGS with C-ToF depth (\cref{eq:tof_depth}) fails to reconstruct dynamic objects well.
    Baseline 2D Flowed model is fast but cannot handle axial motion as well as our model, producing artifacts. 
    \textit{Inset on ground truth: corresponding RGB image. See our supplement for larger images.}
    }
    \label{fig:real-synthetic-baselines}
    \vspace{-0.08cm}
\end{figure*}
\begin{figure*}
    \centering
    \footnotesize

    \newcommand{\imgw}{0.15\linewidth}
    \newcommand{\imgwhalf}{0.075\linewidth}

    \newcommand{\zoomxBaseball}{-0.4} 
    \newcommand{\zoomyBaseball}{-0.45} 

    \newcommand{\zoomxPillow}{-1.05} 
    \newcommand{\zoomyPillow}{0.4} 

    \newcommand{\zoomxTarget}{-0.1} 
    \newcommand{\zoomyTarget}{-0.1} 

    \newcommand{\zoomxFan}{-0.25} 
    \newcommand{\zoomyFan}{0.25} 
    
    \newcommand{\imgDepth}[2][0px 0px 0px 0px]{\includegraphics[draft=\isdraft,width=\imgw,trim=#1,clip,viewport=50 0 290 240]{#2}}
    \newcommand{\zoomDepth}[4]{
        \trimbox{0cm 0cm 0.15cm 0.0cm}{
        \begin{tikzpicture}[
    		image/.style={inner sep=0pt, outer sep=0pt},
    		collabel/.style={above=9pt, anchor=north, inner ysep=0pt, align=center}, 
    		rowlabel/.style={left=9pt, rotate=90, anchor=north, inner ysep=0pt, scale=0.8, align=center},
    		subcaption/.style={inner xsep=0.75mm, inner ysep=0.75mm, below right},
    		arrow/.style={-{Latex[length=2.5mm,width=4mm]}, line width=2mm},
    		spy using outlines={rectangle, size=1.25cm, magnification=2.5, connect spies, ultra thick, every spy on node/.append style={thick}},
    		style1/.style={cyan!90!black,thick},
    		style2/.style={orange!90!black},
    		style3/.style={blue!90!black},
    		style4/.style={green!90!black},
    		style5/.style={white},
    		style6/.style={black},
        ]
        
        \node [image] (#1) {\imgDepth[0px 0px 0px 0px]{#2}};
        \spy[style1] on ($(#1.center)-#3$) in node (crop-#1) [anchor=#4] at (#1.#4);
        
        \end{tikzpicture}
        }
    }

    \vspace{-3mm}
    \setlength{\tabcolsep}{0.5pt}
    \begin{tabular}{ccccccc}
    &
    C-ToF &
    2D Flowed &
    \torf \cite{attal2021torf} &
    \ftorf \cite{okunev2024flowed} &
    DeformableGS \cite{yang2024deformable} &
    Ours
    \\[0.2em]
    
    {\raisebox{0.5in}{\rotatebox[origin=c]{90}{\SeqBaseball}}} &
    \zoomDepth{label1}{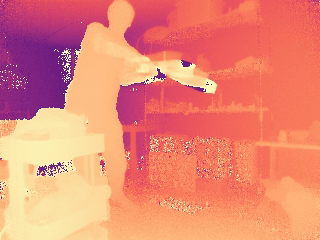}{(\zoomxBaseball,\zoomyBaseball)}{south east} &
    \zoomDepth{label2}{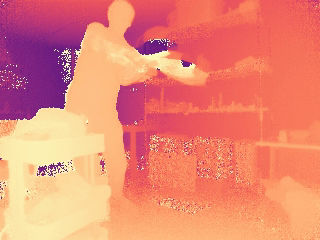}{(\zoomxBaseball,\zoomyBaseball)}{south east} &
    \zoomDepth{label3}{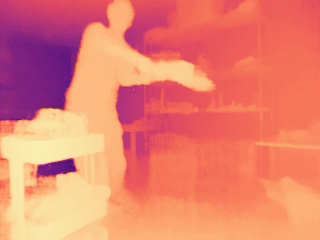}{(\zoomxBaseball,\zoomyBaseball)}{south east} &
    \zoomDepth{label4}{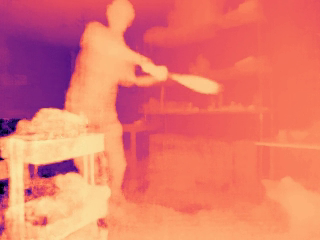}{(\zoomxBaseball,\zoomyBaseball)}{south east} &
    \zoomDepth{label5}{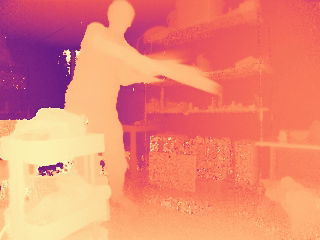}{(\zoomxBaseball,\zoomyBaseball)}{south east} &
    \zoomDepth{label5}{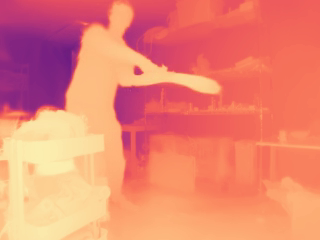}{(\zoomxBaseball,\zoomyBaseball)}{south east} \\

    {\raisebox{0.5in}{\rotatebox[origin=c]{90}{\SeqPillow}}} &
    \zoomDepth{label1}{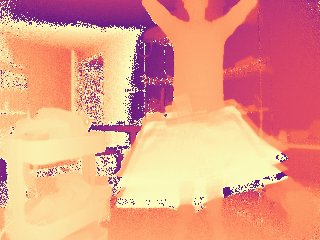}{(\zoomxPillow,\zoomyPillow)}{south west} &
    \zoomDepth{label2}{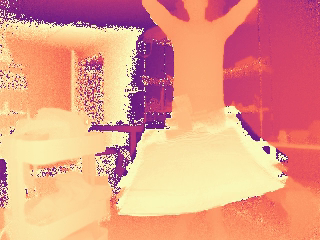}{(\zoomxPillow,\zoomyPillow)}{south west} &
    \zoomDepth{label3}{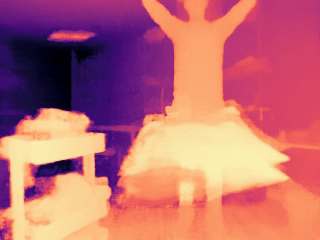}{(\zoomxPillow,\zoomyPillow)}{south west} &
    \zoomDepth{label4}{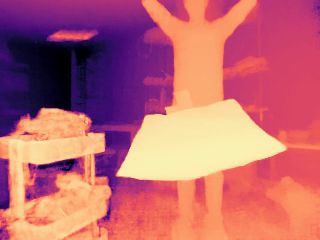}{(\zoomxPillow,\zoomyPillow)}{south west} &
    \zoomDepth{label5}{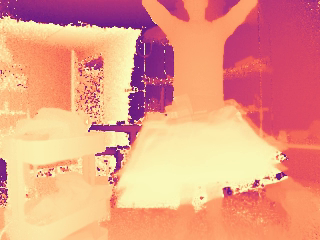}{(\zoomxPillow,\zoomyPillow)}{south west} &
    \zoomDepth{label5}{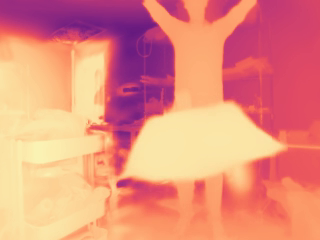}{(\zoomxPillow,\zoomyPillow)}{south west} \\
    
    {\raisebox{0.5in}{\rotatebox[origin=c]{90}{\SeqTarget}}} &
    \zoomDepth{label1}{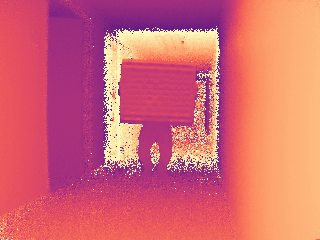}{(\zoomxTarget,\zoomyTarget)}{north west} &
    \zoomDepth{label2}{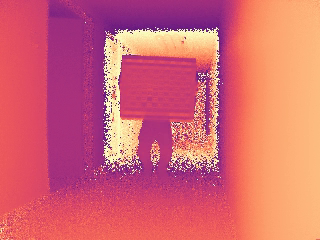}{(\zoomxTarget,\zoomyTarget)}{north west} &
    \zoomDepth{label3}{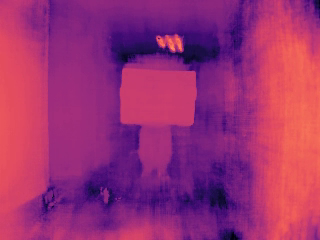}{(\zoomxTarget,\zoomyTarget)}{north west} &
    \zoomDepth{label4}{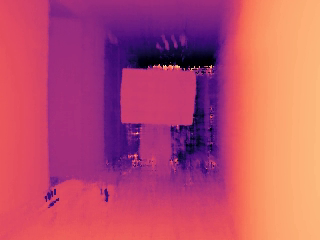}{(\zoomxTarget,\zoomyTarget)}{north west} &
    \zoomDepth{label5}{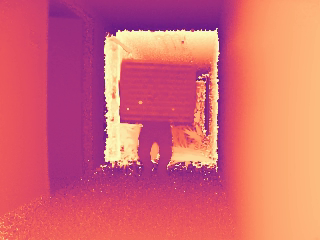}{(\zoomxTarget,\zoomyTarget)}{north west} &
    \zoomDepth{label5}{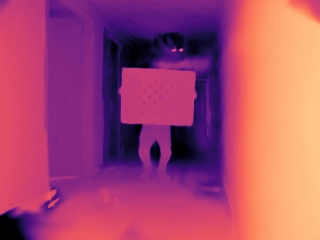}{(\zoomxTarget,\zoomyTarget)}{north west}  \\
    
    {\raisebox{0.5in}{\rotatebox[origin=c]{90}{\SeqFan}}} &
    \zoomDepth{label1}{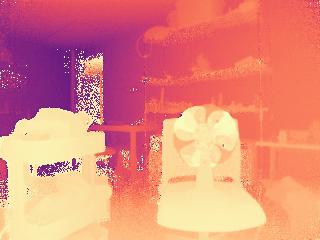}{(\zoomxFan,\zoomyFan)}{south west} &
    \zoomDepth{label2}{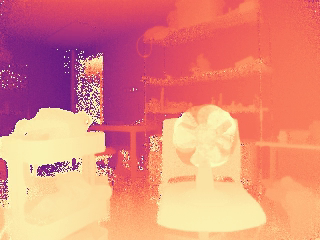}{(\zoomxFan,\zoomyFan)}{south west} &
    \zoomDepth{label3}{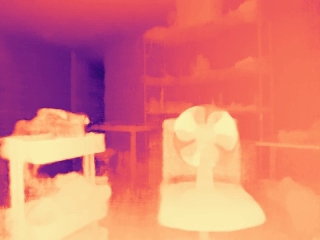}{(\zoomxFan,\zoomyFan)}{south west} &
    \zoomDepth{label4}{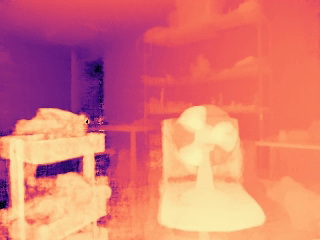}{(\zoomxFan,\zoomyFan)}{south west} &
    \zoomDepth{label5}{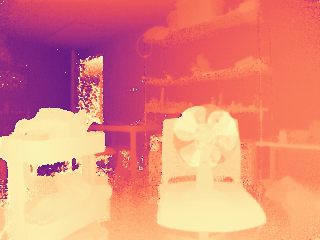}{(\zoomxFan,\zoomyFan)}{south west} &
    \zoomDepth{label5}{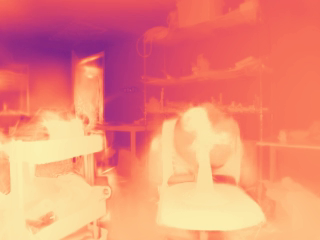}{(\zoomxFan,\zoomyFan)}{south west} \\

    \end{tabular}
    \vspace{-3mm}
    \caption{\textbf{\ftorf real scenes; rendered scene depth $d$.} As it models temporal dynamics and constrains geometry appropriately during training, our model produces comparable quality reconstructions with \ftorf while being consistently better than other baselines. Our method tends to reconstruct static geometry better (floor in \SeqJumpingJacks). \SeqFan scene presents a significant challenge as the motion is nonlinear---our canonical field struggles to track this geometry over time. \ftorf can approximate this scene better due to the lack of such a global constraint. Depth wrapping effects in the background occur in some scenes; these represent an ambiguity that is out of our scope for this work.}
    \label{fig:real-qualitative-baselines}
\end{figure*}
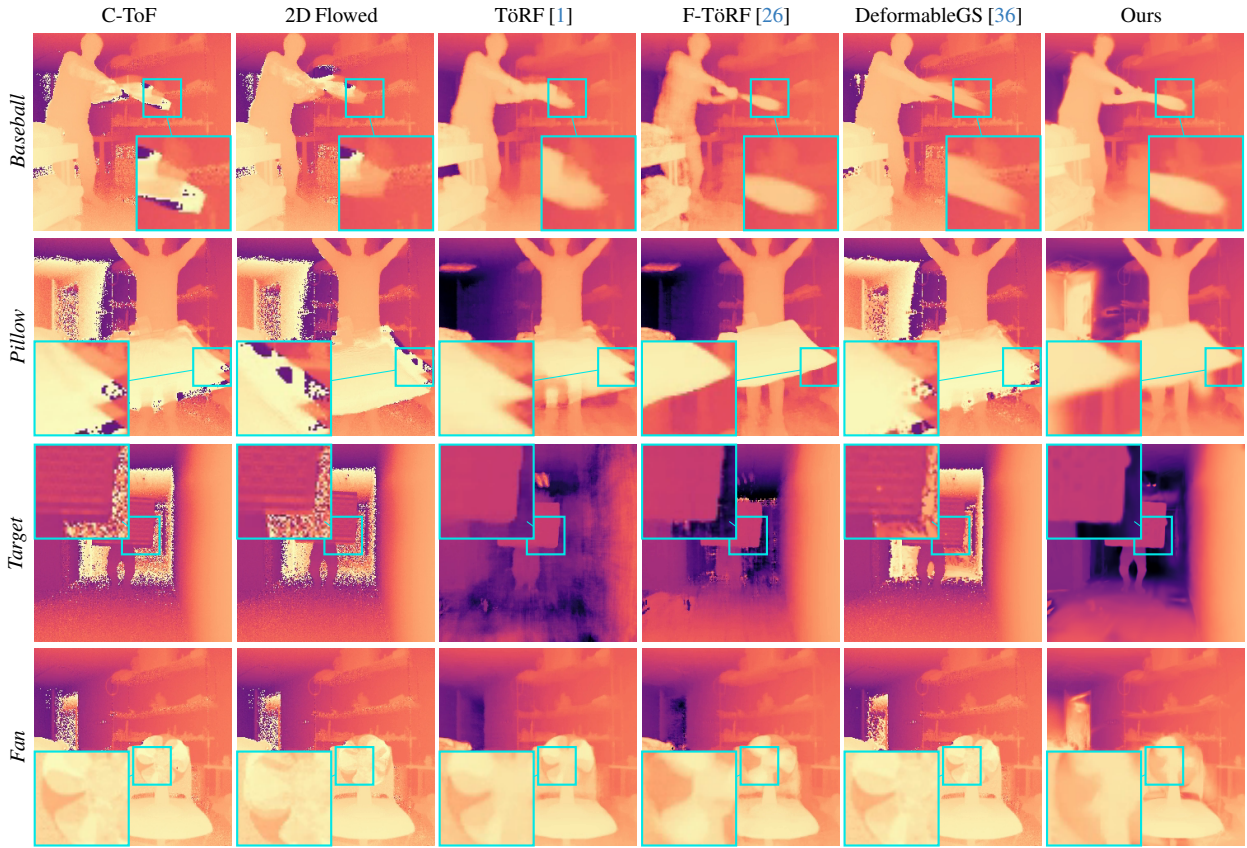
\begin{figure*}
    \centering
    \footnotesize

    \newcommand{\imgw}{0.15\linewidth}

    \newcommand{\zoomxCupboard}{-0.31} 
    \newcommand{\zoomyCupboard}{-0.52} 

    \newcommand{\zoomxStudybook}{0.1} 
    \newcommand{\zoomyStudybook}{-0.2} 

    \newcommand{\imgDepth}[2][0px 0px 0px 0px]{\includegraphics[draft=\isdraft,width=\imgw,trim=#1,clip,viewport=50 0 290 240]{#2}}
    \newcommand{\zoomDepth}[4]{
        \trimbox{0cm 0cm 0.15cm 0.0cm}{
        \begin{tikzpicture}[
    		image/.style={inner sep=0pt, outer sep=0pt},
    		collabel/.style={above=9pt, anchor=north, inner ysep=0pt, align=center}, 
    		rowlabel/.style={left=9pt, rotate=90, anchor=north, inner ysep=0pt, scale=0.8, align=center},
    		subcaption/.style={inner xsep=0.75mm, inner ysep=0.75mm, below right},
    		arrow/.style={-{Latex[length=2.5mm,width=4mm]}, line width=2mm},
    		spy using outlines={rectangle, size=1.25cm, magnification=2.5, connect spies, ultra thick, every spy on node/.append style={thick}},
    		style1/.style={cyan!90!black,thick},
    		style2/.style={orange!90!black},
    		style3/.style={blue!90!black},
    		style4/.style={green!90!black},
    		style5/.style={white},
    		style6/.style={black},
        ]
        
        \node [image] (#1) {\imgDepth[0px 0px 0px 0px]{#2}};
        \spy[style1] on ($(#1.center)-#3$) in node (crop-#1) [anchor=#4] at (#1.#4);
        
        \end{tikzpicture}
        }
    }

    \newcommand{\zoomDepthCupboard}[4]{
        \trimbox{0cm 0cm 0.15cm 0.0cm}{
        \begin{tikzpicture}[
    		image/.style={inner sep=0pt, outer sep=0pt},
    		collabel/.style={above=9pt, anchor=north, inner ysep=0pt, align=center}, 
    		rowlabel/.style={left=9pt, rotate=90, anchor=north, inner ysep=0pt, scale=0.8, align=center},
    		subcaption/.style={inner xsep=0.75mm, inner ysep=0.75mm, below right},
    		arrow/.style={-{Latex[length=2.5mm,width=4mm]}, line width=2mm},
    		spy using outlines={rectangle, size=1.25cm, magnification=1.8, connect spies, ultra thick, every spy on node/.append style={thick}},
    		style1/.style={cyan!90!black,thick},
    		style2/.style={orange!90!black},
    		style3/.style={blue!90!black},
    		style4/.style={green!90!black},
    		style5/.style={white},
    		style6/.style={black},
        ]
        
        \node [image] (#1) {\imgDepth[0px 0px 0px 0px]{#2}};
        \spy[style1] on ($(#1.center)-#3$) in node (crop-#1) [anchor=#4] at (#1.#4);
        
        \end{tikzpicture}
        }
    }

    \vspace{-3mm}
    \setlength{\tabcolsep}{0.5pt}
    \begin{tabular}{cccccc}
    &
    Input &
    Ours &
    TöRF \cite{attal2021torf} &
    DeformableGS \cite{yang2024deformable} &
    DeformableGS (--depth) 
    \\
    
    \multirow{2}*{\raisebox{0.0in}{\rotatebox[origin=c]{90}{\SeqCupboard}}} &
    \zoomDepthCupboard{label1}{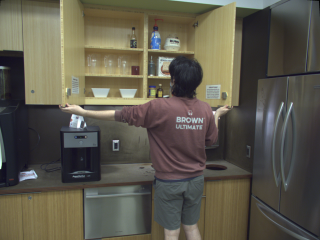}{(\zoomxCupboard+0.2,\zoomyCupboard+0.1)}{south west} &
    \zoomDepthCupboard{label2}{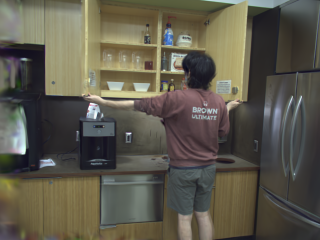}{(\zoomxCupboard,\zoomyCupboard)}{south west} &
    \zoomDepthCupboard{label3}{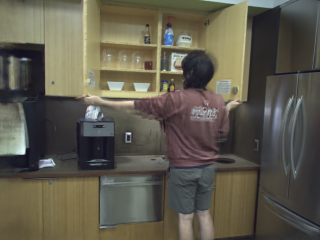}{(\zoomxCupboard,\zoomyCupboard)}{south west} &
    \zoomDepthCupboard{label4}{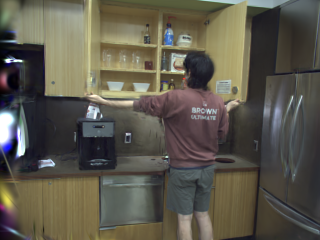}{(\zoomxCupboard,\zoomyCupboard)}{south west} &
    \zoomDepthCupboard{label5}{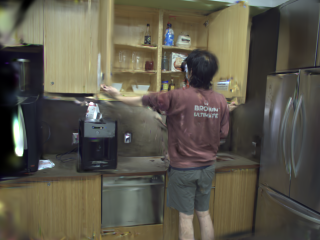}{(\zoomxCupboard,\zoomyCupboard)}{south west} \\
    &
    \zoomDepthCupboard{label1}{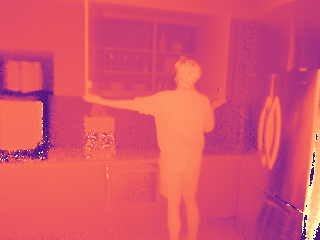}{(\zoomxCupboard+0.15,\zoomyCupboard+0.0)}{south west} &
    \zoomDepthCupboard{label2}{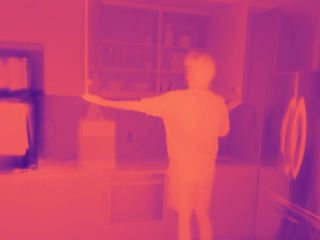}{(\zoomxCupboard,\zoomyCupboard)}{south west} &
    \zoomDepthCupboard{label3}{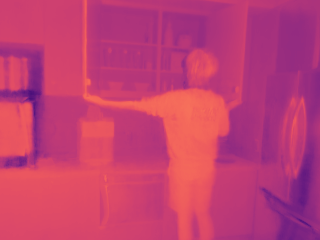}{(\zoomxCupboard,\zoomyCupboard)}{south west} &
    \zoomDepthCupboard{label4}{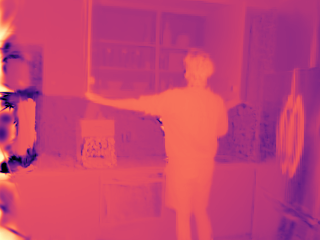}{(\zoomxCupboard,\zoomyCupboard)}{south west} &
    \zoomDepthCupboard{label5}{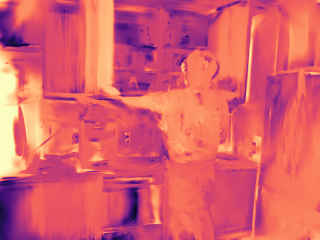}{(\zoomxCupboard,\zoomyCupboard)}{south west} \\
    
    \multirow{2}{*}{\raisebox{0.0\height}{\rotatebox[origin=c]{90}{\SeqStudyBook}}} &
    \zoomDepth{label1}{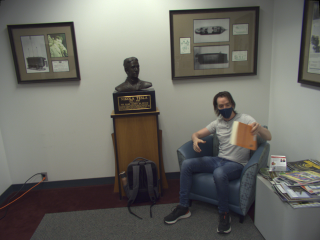}{(\zoomxStudybook+0.18,\zoomyStudybook-0.05)}{north east} &
    \zoomDepth{label2}{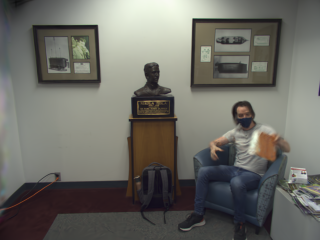}{(\zoomxStudybook,\zoomyStudybook)}{north east} &
    \zoomDepth{label3}{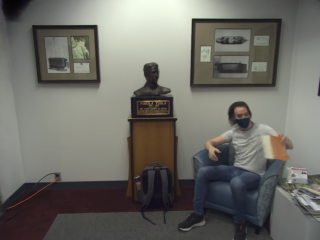}{(\zoomxStudybook,\zoomyStudybook)}{north east} &
    \zoomDepth{label4}{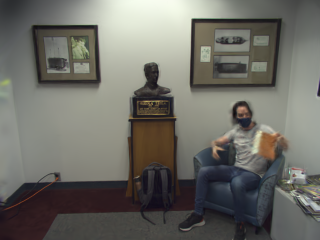}{(\zoomxStudybook,\zoomyStudybook)}{north east} &
    \zoomDepth{label5}{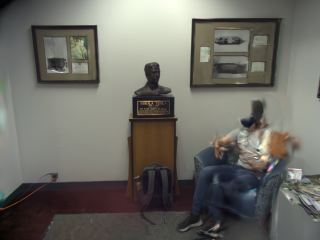}{(\zoomxStudybook,\zoomyStudybook)}{north east} \\
    &
    \zoomDepth{label1}{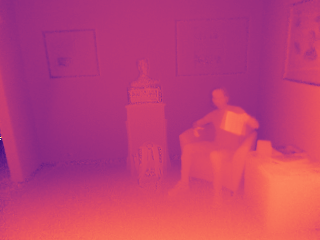}{(\zoomxStudybook+0.13,\zoomyStudybook-0.18)}{north east} &
    \zoomDepth{label2}{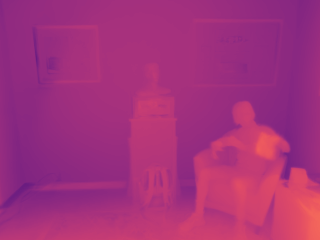}{(\zoomxStudybook,\zoomyStudybook)}{north east} &
    \zoomDepth{label3}{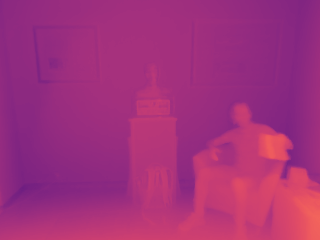}{(\zoomxStudybook,\zoomyStudybook)}{north east} &
    \zoomDepth{label4}{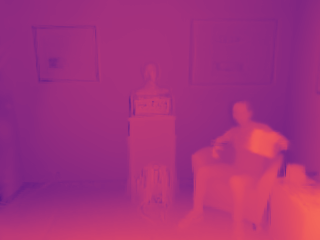}{(\zoomxStudybook,\zoomyStudybook)}{north east} &
    \zoomDepth{label5}{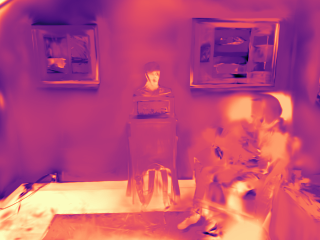}{(\zoomxStudybook,\zoomyStudybook)}{north east} \\

    \end{tabular}
    \vspace{-3mm}
    \caption{\textbf{Our approach is competitive with other baselines on \torf dataset.} Comparison of our method with TöRF and DeformableGS (with and without C-ToF depth prior). Images are novel views of rendered volumetric depth $d$ and color, rendered along a spiral path around the training camera. Our approach demonstrates same or better depth reconstruction for both static and dynamic objects, particularly in dynamic regions with limited multi-view signals. Our method mitigates overfitting to noisy raw depth measurements in the \SeqStudyBook scene.}
    \label{fig:torf-qualitative-baselines}
\end{figure*}

\vspace{-5pt}
\inlineheading{\torf dataset}
The dataset contains five real-world indoor dynamic monocular sequences (\SeqPhotocopier, \SeqCupboard, \SeqDeskBox, \SeqPhoneBooth, and \SeqStudyBook). All scenes have raw C-ToF and RGB images with 12 bits of information per pixel. C-ToF signals were captured with a Texas Instruments OPT8241 sensor (320$\times$240 at 30\,fps) with 5\,m unambiguous depth range. We use the camera poses optimized by the \torf model. The cameras are moving, which provides additional multi-view constraints, and scene motion is slow.

\inlineheading{\ftorf dataset}
The dataset contains five sequences from the dataset: \SeqPillow, \SeqBaseball, \SeqJumpingJacks, \SeqTarget, and \SeqFan. Again, these scenes were captured with a Texas Instruments OPT8241 sensor (320$\times$240 at 30\,fps) with 5\,m unambiguous depth range. The scenes are significantly more challenging: they contain no camera motion and they contain fast-moving objects such as falling pillows and swinging baseball bats.

\inlineheading{\ftorf synthetic dataset}
We also use the seven synthetic scenes generated in Blender using a physically-based path tracer PBRT \cite{pharr2016physically} adapted for C-ToF emission, at a resolution of 320$\times$240. The scenes show cubes undergoing axial, lateral, and rotational motions to induce varying disparities, with occlusion, both with and without texture, or with chairs to show thin features.
The scenes are strictly monocular without camera motion.
The fastest scenes (\SeqThreeCubes, \SeqThreeChairs, \SeqArcingCube, \SeqAxial) have large disparity: maximally up to 30 pixels across raw frames, and often 9--18 pixels (\SeqSlidingCube, \SeqOccludedCube, \SeqZMotion). Three challenging scenes test large axial motion (\SeqArcingCube, \SeqAxial, \SeqZMotion).

\inlineheading{Metrics}
For synthetic scenes, we report depth MSE from the ground truth. For real-world scenes, we only evaluate qualitatively. We report compute times for an NVIDIA 3090 GPU.

\inlineheading{Baselines}
We compare to the phasor-based C-ToF method \torf (\citet{attal2021torf}) that assumes synchronous capture, and to \ftorf (\citet{okunev2024flowed}) that directly uses asynchronous raw C-ToF measurements. Both are neural volume based. We also compare to DeformableGS (\citet{yang2024deformable}), a dynamic Gaussian splatting reconstruction method, where we add a loss between the C-ToF depth from \cref{eq:tof_depth} and the scene depth from \cref{eq:Depth_Image_Formation}. Lastly, we compare to a 2D baseline that uses softmax splatting (\citet{niklaus2020softmax}) and RAFT optical flow to warp raw quads to new timesteps.

\subsection{Results}
\vspace{-5pt}
\inlineheading{Computation time}
As expected, our approach is significantly faster than \torf~\cite{attal2021torf} (60 hours) and \ftorf~\cite{okunev2024flowed} (72 hours) at 40--60 minutes per 4D scene reconstruction. Rendering is real time at 100+\,Hz.

\inlineheading{Quantitative results}
Our method is competitive in terms of depth error (\cref{quantitative-synthetic-table}). For the reconstructed scene depth $d$, we show improved results on five out of seven synthetic scenes over all baselines and produce competitive results on the rest. Our predicted scene density is `denser' than that of \ftorf (\SeqArcingCube), as seen by the reduced gap between $d$ and the depth $d_\text{ToF}$ derived from the reconstructed sensor images using \cref{eq:tof_depth}. Fast-moving thin structures (as in \SeqThreeChairs) seem harder to reconstruct as they require a precise configuration of anisotropic Gaussians. Large-scale nonlinear motions (\SeqArcingCube) are also challenging to optimize under a canonical deformation model. \SeqZMotion suffers from a local optimum on one of the cubes, which biases the error metric upward (see the supplemental videos).

\inlineheading{Qualitative results}
On the easier \torf dataset that has color and a moving camera (\cref{fig:torf-qualitative-baselines}), we show that depth supervision is still critical: DeformableGS without the added depth loss fails to reconstruct the scene well. With C-ToF depth, DeformableGS improves substantially, but is still inferior to our physically-based C-ToF supervision. Our method is comparable to \torf, with improvements over some areas (e.g., dark hair region in \SeqCupboard). 

The challenging \ftorf dataset (\cref{fig:real-synthetic-baselines,fig:real-qualitative-baselines}) reveals that our method is comparable with \ftorf, and only these two methods can correct for fast motion artifacts. DeformableGS often completely fails to reconstruct a fast-moving object (\SeqArcingCube, \SeqThreeChairs), and 2D Flowed cannot handle axial motion---something that our method can do. While faster, our method also often produces better static region reconstructions compared to \ftorf (floor in \SeqJumpingJacks, \SeqTarget, cart in \SeqBaseball). But, similar to the quantitative analysis, our method somewhat struggles with fast-moving thin structures in \SeqThreeChairs and can produce depth that is less sharp than \ftorf. 
Finally, no approach can accurately reconstruct spinning fan blades in \SeqFan.
\vspace{0.05cm}
\section{Discussion}
\vspace{-3pt}
\inlineheading{Limitations}
\textit{Depth wrapping.}
Several of our reconstructions (\SeqJumpingJacks, \SeqTarget, \SeqPillow) contain depth wrapping in the background, which is a typical C-ToF artifact. Although some baselines happen to avoid this issue, this problem is fundamentally ill-posed for a static camera, and predictions heavily depend upon the optimization biases since the true solution is ambiguous.

\textit{Nonlinear motion.}
Our model can struggle when the motion is strongly nonlinear (as in \SeqFan and to a smaller extent in \SeqArcingCube). This can, in principle, be resolved with a more flexible model that allows non-piecewise-linear trajectories, although it will likely make the problem even more ill-posed and require stronger regularization.

\textit{Sharp reconstruction.}
Our volume reconstruction can be visually less sharp than NeRF-based methods around the edges, despite having smaller depth error. This is likely a consequence of the spatial smoothness that Gaussians exhibit. More complex surface representations might help \cite{guedon2023sugar, huang20242d}.

\textit{Thin and fast-moving objects.} Our deformation MLP might not maintain thin object structures during large deformation. Since the loss is computed in deformed space but Gaussians are densified in canonical space, this causes unstable densification and makes thin \emph{and} fast-moving objects harder to reconstruct. Longer motion tracks or spacetime rigidity might help \cite{stearns2024marbles}, and this is left for future work.

\textit{No universal guarantees.} Our heuristics are based on empirical observations and may not be suitable for different scenes.

\textit{Limited indoor scenes.} C-ToF imaging may struggle in outdoor scenes, and many motions exist in the real world that go beyond our set of planar, arcing, and axial motions.

\vspace{-0.1cm}
\section{Related Works}
\vspace{-0.1cm}

Accurately reconstructing geometry is important for many applications, such as in measurement or physics simulation \cite{xie2024physgaussian}.
When the capturing setup is equipped with a depth sensor, high-quality reconstruction becomes possible even with a small camera baseline.
In \torf \cite{attal2021torf}, a NeRF model takes advantage of a C-ToF camera and uses phasors to optimize the geometry of a dynamic scene.
In \ftorf \cite{okunev2024flowed} the model works directly with raw C-ToF camera outputs and recovers motion in addition to geometry, while avoiding typical C-ToF camera motion artifacts.
Our work adapts such techniques to fast Gaussian splatting.
Neural structured light approaches can also recover geometry, normals, and direct and indirect illumination \cite{shandilya2023neural,Mirdehghan2024CVPR}.
For lidar, PlatoNeRF \cite{PlatoNeRF} recovers scene geometry from a single view using two-bounce signals captured by a single-photon lidar.
Snapshot Lidar \cite{friday2024snapshot} and \citet{lee2024light} improve C-ToF temporal resolution but they require hardware-level control.

\inlineheading{Dynamic Gaussian splatting} 
Since 3DGS \cite{kerbl20233d} was introduced, it has been explored and adapted widely for reconstruction due to its efficiency. 
Extending 3DGS, a plethora of works on dynamic Gaussian splatting emerged recently, differing in their motion representations, constraints and optimization approaches.
In Dynamic 3D Gaussians \cite{luiten2023dynamic}, the model assumes a freeform rotation and translation of Gaussians between consecutive time steps and iteratively optimizes the scene with rigidity constraints, but it heavily relies on multi-view input.
Some works \cite{yang2024deformable, wu20244d, liang2023gaufre} map each time moment to a canonical space to model the dynamic content.
Another approach to model the motion is a global trajectory reconstruction using Fourier, polynomial or a learned basis approximation \cite{katsumata2023efficient, kratimenos2024dynmf, lin2023gaussian, som2024}.
In Dynamic Gaussian Marbles \cite{stearns2024marbles}, isotropic Gaussians equipped with trajectories are learned through a divide-and-conquer optimization using a pretrained single-image depth prior.
A local approach is taken in Spacetime Gaussians \cite{li2023spacetime}: the temporal opacity of Gaussians is modeled through radial basis functions making them appear and disappear when needed.
Finally, 4D Gaussian splatting \cite{yang2023gs4d} avoids using an explicit motion model at all by using 4D Gaussians, where the motion is a byproduct of time-conditioned projection to 3D space.

\inlineheading{Gaussian splatting for other input domains}
Gaussian splatting models are not restricted to only reconstruct from RGB color inputs.
The method can be adapted to a non-RGB setting as well.
Some works \cite{li2024chaos, jin2024lighting, qu2024z, cai2024radiative, chen2024thermal3d} focus on extending 3DGS framework to HDR, X-ray, sonar, or thermal images.
However, without a sufficient variation in camera poses, high quality reconstruction remains a challenge.
To address that, some methods \cite{chung2024depth} integrate an additional depth input to improve the quality.
However, these methods all rely on the quality of the depth priors.
Within an under-constrained and more brittle optimization than NeRFs, we contribute a way to effectively use Gaussian splatting with C-ToF imaging.

\vspace{-0.1cm}
\section{Conclusion}
We present a Gaussian splatting approach for fast and accurate reconstruction of dynamic monocular sequences with asynchronous C-ToF exposures. Adapting GS to this setting required us to consider how the underlying optimization affects the indirect estimation of depth, and devise two heuristics that better condition the optimization. These successfully avoid overfitting across a set of synthetic and real-world scenes. In sum, our method is 100$\times$ faster than baselines, and produces comparable or better reconstructions of fast-moving objects even with a static monocular camera.

\vspace{0.1cm}
\inlineheading{Acknowledgements}
RL, MO, ZG, AHD, JT thank NSF CAREER 2144956, NASA RI-80NSSC23M0075, and Cognex. MOT thanks NSF CAREER 2238485.

{
    \small
    \bibliographystyle{ieeenat_fullname}
    \bibliography{main}
}

\appendix
\newcommand{\maineq}[1]{Eq.~(#1) in the main paper}
\newcommand{\mainfig}[1]{Fig.~(#1) in the main paper}

\clearpage
\maketitlesupplementary

\section{Supplemental Video + Full Figures}

We have included a supplemental video showing all dataset sequences plus two ablation sequences. Further, we include full sequence qualitative figures for all three datasets at the back of this supplemental material; this reproduces some results from the main paper but groups and includes all results for completeness.

\section{Ablations}

\begin{figure*}[t]
    \centering
    \footnotesize
    \vspace{-5mm}

    \newcommand{\imgw}{0.15\linewidth}

    \newcommand{\zoomxAcuteZSpeedTest}{-0.2} 
    \newcommand{\zoomyAcuteZSpeedTest}{-0.3} 

    \newcommand{\zoomxArcingCube}{0.1} 
    \newcommand{\zoomyArcingCube}{-0.45} 

    \newcommand{\zoomxSpeedTestChair}{-0.15} 
    \newcommand{\zoomySpeedTestChair}{-0.2} 

    \newcommand{\zoomxBaseball}{0.65} 
    \newcommand{\zoomyBaseball}{-0.65} 

    \newcommand{\zoomxPillow}{0.95} 
    \newcommand{\zoomyPillow}{0.7} 

    \newcommand{\colorbarTrimLeft}{-0.5}
    \newcommand{\colorbarTrimTop}{0.05}

    \newcommand{\imgDepth}[2][0px 0px 0px 0px]{\includegraphics[draft=\isdraft,width=\imgw,trim=#1,clip,viewport=0 0 320 240]{#2}}
    \newcommand{\zoomDepth}[4]{
        \begin{tikzpicture}[
    		image/.style={inner sep=0pt, outer sep=0pt},
    		collabel/.style={above=9pt, anchor=north, inner ysep=0pt, align=center}, 
    		rowlabel/.style={left=9pt, rotate=90, anchor=north, inner ysep=0pt, scale=0.8, align=center},
    		subcaption/.style={inner xsep=0.75mm, inner ysep=0.75mm, below right},
    		arrow/.style={-{Latex[length=2.5mm,width=4mm]}, line width=2mm},
    		spy using outlines={rectangle, size=1.00cm, magnification=2.5, connect spies, ultra thick, every spy on node/.append style={thick}},
    		style1/.style={cyan!90!black,thick},
    		style2/.style={orange!90!black},
    		style3/.style={blue!90!black},
    		style4/.style={green!90!black},
    		style5/.style={white},
    		style6/.style={black},
        ]
        
        \node [image] (#1) {\imgDepth[0px 0px 0px 0px]{#2}};
        \spy[style1] on ($(#1.center)-#3$) in node (crop-#1) [anchor=#4] at (#1.#4);
        
        \end{tikzpicture}
    }

    \newcommand{\zoomDepthImgOnly}[4]{
        \begin{tikzpicture}[
    		image/.style={inner sep=0pt, outer sep=0pt},
    		collabel/.style={above=9pt, anchor=north, inner ysep=0pt, align=center}, 
    		rowlabel/.style={left=9pt, rotate=90, anchor=north, inner ysep=0pt, scale=0.8, align=center},
    		subcaption/.style={inner xsep=0.75mm, inner ysep=0.75mm, below right},
    		arrow/.style={-{Latex[length=2.5mm,width=4mm]}, line width=2mm},
    		spy using outlines={rectangle, size=1.25cm, magnification=2.5, connect spies, ultra thick, every spy on node/.append style={thick}},
    		style1/.style={cyan!90!black,thick},
    		style2/.style={orange!90!black},
    		style3/.style={blue!90!black},
    		style4/.style={green!90!black},
    		style5/.style={white},
    		style6/.style={black},
        ]
        
        \node [image] (#1) {\imgDepth[0px 0px 0px 0px]{#2}};
        
        \end{tikzpicture}
    }
    
    \setlength{\tabcolsep}{0.5pt}
    \begin{tabular}{cccccccc}
        & & Input & No bias & +DD & +H1 & +H1 +H2 (ours) & +H1 +H2 +DD \\
        \multirow{2}*{\raisebox{0.0in}{\rotatebox[origin=c]{90}{\SeqZMotionShort}}} &
        {\raisebox{0.3in}{\rotatebox[origin=c]{90}{$d_\text{{ToF}}$\xspace}}} &
        \zoomDepthImgOnly{label}{images/ablations/acute_z_speed_test/input\_depth.png}{(\zoomxAcuteZSpeedTest, \zoomyAcuteZSpeedTest)}{north west} &
        \zoomDepthImgOnly{label}{images/ablations/acute_z_speed_test/no\_bias\_depth_tof.png}{(\zoomxAcuteZSpeedTest, \zoomyAcuteZSpeedTest)}{north west} &
        \zoomDepthImgOnly{label}{images/ablations/acute_z_speed_test/+dd\_depth_tof.png}{(\zoomxAcuteZSpeedTest, \zoomyAcuteZSpeedTest)}{north west} &
        \zoomDepthImgOnly{label}{images/ablations/acute_z_speed_test/+h1\_depth_tof.png}{(\zoomxAcuteZSpeedTest, \zoomyAcuteZSpeedTest)}{north west} &
        \zoomDepthImgOnly{label}{images/ablations/acute_z_speed_test/+h1\_+h2\_depth_tof.png}{(\zoomxAcuteZSpeedTest, \zoomyAcuteZSpeedTest)}{north west} &
        \zoomDepthImgOnly{label}{images/ablations/acute_z_speed_test/+h1\_+h2\_+dd\_depth_tof.png}{(\zoomxAcuteZSpeedTest, \zoomyAcuteZSpeedTest)}{north west} \\
        & {\raisebox{0.4in}{\rotatebox[origin=c]{90}{$d$\xspace}}} &
        \zoomDepthImgOnly{label}{images/ablations/acute_z_speed_test/input\_reflectivity.png}{(\zoomxAcuteZSpeedTest, \zoomyAcuteZSpeedTest)}{north west} &
        \zoomDepthImgOnly{label}{images/ablations/acute_z_speed_test/no\_bias\_depth.png}{(\zoomxAcuteZSpeedTest, \zoomyAcuteZSpeedTest)}{north west} &
        \zoomDepthImgOnly{label}{images/ablations/acute_z_speed_test/+dd\_depth.png}{(\zoomxAcuteZSpeedTest, \zoomyAcuteZSpeedTest)}{north west} &
        \zoomDepthImgOnly{label}{images/ablations/acute_z_speed_test/+h1\_depth.png}{(\zoomxAcuteZSpeedTest, \zoomyAcuteZSpeedTest)}{north west} &
        \zoomDepthImgOnly{label}{images/ablations/acute_z_speed_test/+h1\_+h2\_depth.png}{(\zoomxAcuteZSpeedTest, \zoomyAcuteZSpeedTest)}{north west} &
        \zoomDepthImgOnly{label}{images/ablations/acute_z_speed_test/+h1\_+h2\_+dd\_depth.png}{(\zoomxAcuteZSpeedTest, \zoomyAcuteZSpeedTest)}{north west} \\
        \multirow{2}*{\raisebox{0.0in}{\rotatebox[origin=c]{90}{\SeqThreeChairsShort}}} &
        {\raisebox{0.3in}{\rotatebox[origin=c]{90}{$d_\text{{ToF}}$\xspace}}} &
        \zoomDepth{label}{images/ablations/speed_test_chair/input\_depth.png}{(\zoomxSpeedTestChair, \zoomySpeedTestChair)}{south east} &
        \zoomDepth{label}{images/ablations/speed_test_chair/no\_bias\_depth_tof.png}{(\zoomxSpeedTestChair, \zoomySpeedTestChair)}{south east} &
        \zoomDepth{label}{images/ablations/speed_test_chair/+dd\_depth_tof.png}{(\zoomxSpeedTestChair, \zoomySpeedTestChair)}{south east} &
        \zoomDepth{label}{images/ablations/speed_test_chair/+h1\_depth_tof.png}{(\zoomxSpeedTestChair, \zoomySpeedTestChair)}{south east} &
        \zoomDepth{label}{images/ablations/speed_test_chair/+h1\_+h2\_depth_tof.png}{(\zoomxSpeedTestChair, \zoomySpeedTestChair)}{south east} &
        \zoomDepth{label}{images/ablations/speed_test_chair/+h1\_+h2\_+dd\_depth_tof.png}{(\zoomxSpeedTestChair, \zoomySpeedTestChair)}{south east} \\
        & {\raisebox{0.4in}{\rotatebox[origin=c]{90}{$d$\xspace}}} &
        \zoomDepth{label}{images/ablations/speed_test_chair/input\_reflectivity.png}{(\zoomxSpeedTestChair, \zoomySpeedTestChair)}{south east} &
        \zoomDepth{label}{images/ablations/speed_test_chair/no\_bias\_depth.png}{(\zoomxSpeedTestChair, \zoomySpeedTestChair)}{south east} &
        \zoomDepth{label}{images/ablations/speed_test_chair/+dd\_depth.png}{(\zoomxSpeedTestChair, \zoomySpeedTestChair)}{south east} &
        \zoomDepth{label}{images/ablations/speed_test_chair/+h1\_depth.png}{(\zoomxSpeedTestChair, \zoomySpeedTestChair)}{south east} &
        \zoomDepth{label}{images/ablations/speed_test_chair/+h1\_+h2\_depth.png}{(\zoomxSpeedTestChair, \zoomySpeedTestChair)}{south east} &
        \zoomDepth{label}{images/ablations/speed_test_chair/+h1\_+h2\_+dd\_depth.png}{(\zoomxSpeedTestChair, \zoomySpeedTestChair)}{south east} \\
        \multirow{2}*{\raisebox{0.0in}{\rotatebox[origin=c]{90}{\SeqArcingCube}}} &
        {\raisebox{0.3in}{\rotatebox[origin=c]{90}{$d_\text{{ToF}}$\xspace}}} &
        \zoomDepth{label}{images/ablations/arcing_cube/input\_depth.png}{(\zoomxArcingCube, \zoomyArcingCube)}{south east} &
        \zoomDepth{label}{images/ablations/arcing_cube/no\_bias\_depth_tof.png}{(\zoomxArcingCube, \zoomyArcingCube)}{south east} &
        \zoomDepth{label}{images/ablations/arcing_cube/+dd\_depth_tof.png}{(\zoomxArcingCube, \zoomyArcingCube)}{south east} &
        \zoomDepth{label}{images/ablations/arcing_cube/+h1\_depth_tof.png}{(\zoomxArcingCube, \zoomyArcingCube)}{south east} &
        \zoomDepth{label}{images/ablations/arcing_cube/+h1\_+h2\_depth_tof.png}{(\zoomxArcingCube, \zoomyArcingCube)}{south east} &
        \zoomDepth{label}{images/ablations/arcing_cube/+h1\_+h2\_+dd\_depth_tof.png}{(\zoomxArcingCube, \zoomyArcingCube)}{south east} \\
        & {\raisebox{0.4in}{\rotatebox[origin=c]{90}{$d$\xspace}}} &
        \zoomDepth{label}{images/ablations/arcing_cube/input\_reflectivity.png}{(\zoomxArcingCube, \zoomyArcingCube)}{south east} &
        \zoomDepth{label}{images/ablations/arcing_cube/no\_bias\_depth.png}{(\zoomxArcingCube, \zoomyArcingCube)}{south east} &
        \zoomDepth{label}{images/ablations/arcing_cube/+dd\_depth.png}{(\zoomxArcingCube, \zoomyArcingCube)}{south east} &
        \zoomDepth{label}{images/ablations/arcing_cube/+h1\_depth.png}{(\zoomxArcingCube, \zoomyArcingCube)}{south east} &
        \zoomDepth{label}{images/ablations/arcing_cube/+h1\_+h2\_depth.png}{(\zoomxArcingCube, \zoomyArcingCube)}{south east} &
        \zoomDepth{label}{images/ablations/arcing_cube/+h1\_+h2\_+dd\_depth.png}{(\zoomxArcingCube, \zoomyArcingCube)}{south east} \\
        \arrayrulecolor{black!100}\midrule
        \multirow{2}*{\raisebox{0.0in}{\rotatebox[origin=c]{90}{\SeqPillow}}} &
        {\raisebox{0.3in}{\rotatebox[origin=c]{90}{$d_\text{{ToF}}$\xspace}}} &
        \zoomDepth{label}{images/ablations/pillow/input\_depth.png}{(\zoomxPillow, \zoomyPillow)}{north west} &
        \zoomDepth{label}{images/ablations/pillow/no\_bias\_depth_tof.png}{(\zoomxPillow, \zoomyPillow)}{north west} &
        \zoomDepth{label}{images/ablations/pillow/+dd\_depth_tof.png}{(\zoomxPillow, \zoomyPillow)}{north west} &
        \zoomDepth{label}{images/ablations/pillow/+h1\_depth_tof.png}{(\zoomxPillow, \zoomyPillow)}{north west} &
        \zoomDepth{label}{images/ablations/pillow/+h1\_+h2\_depth_tof.png}{(\zoomxPillow, \zoomyPillow)}{north west} &
        \zoomDepth{label}{images/ablations/pillow/+h1\_+h2\_+dd\_depth_tof.png}{(\zoomxPillow, \zoomyPillow)}{north west} \\
        & {\raisebox{0.4in}{\rotatebox[origin=c]{90}{$d$\xspace}}} &
        \zoomDepth{label}{images/ablations/pillow/input\_reflectivity.png}{(\zoomxPillow, \zoomyPillow)}{north west} &
        \zoomDepth{label}{images/ablations/pillow/no\_bias\_depth.png}{(\zoomxPillow, \zoomyPillow)}{north west} &
        \zoomDepth{label}{images/ablations/pillow/+dd\_depth.png}{(\zoomxPillow, \zoomyPillow)}{north west} &
        \zoomDepth{label}{images/ablations/pillow/+h1\_depth.png}{(\zoomxPillow, \zoomyPillow)}{north west} &
        \zoomDepth{label}{images/ablations/pillow/+h1\_+h2\_depth.png}{(\zoomxPillow, \zoomyPillow)}{north west} &
        \zoomDepth{label}{images/ablations/pillow/+h1\_+h2\_+dd\_depth.png}{(\zoomxPillow, \zoomyPillow)}{north west} \\
        \multirow{2}*{\raisebox{0.0in}{\rotatebox[origin=c]{90}{\SeqBaseball}}} &
        {\raisebox{0.3in}{\rotatebox[origin=c]{90}{$d_\text{{ToF}}$\xspace}}} &
        \zoomDepth{label}{images/ablations/baseball/input\_depth.png}{(\zoomxBaseball, \zoomyBaseball)}{south east} &
        \zoomDepth{label}{images/ablations/baseball/no\_bias\_depth_tof.png}{(\zoomxBaseball, \zoomyBaseball)}{south east} &
        \zoomDepth{label}{images/ablations/baseball/+dd\_depth_tof.png}{(\zoomxBaseball, \zoomyBaseball)}{south east} &
        \zoomDepth{label}{images/ablations/baseball/+h1\_depth_tof.png}{(\zoomxBaseball, \zoomyBaseball)}{south east} &
        \zoomDepth{label}{images/ablations/baseball/+h1\_+h2\_depth_tof.png}{(\zoomxBaseball, \zoomyBaseball)}{south east} &
        \zoomDepth{label}{images/ablations/baseball/+h1\_+h2\_+dd\_depth_tof.png}{(\zoomxBaseball, \zoomyBaseball)}{south east} \\
        & {\raisebox{0.4in}{\rotatebox[origin=c]{90}{$d$\xspace}}} &
        \zoomDepth{label}{images/ablations/baseball/input\_reflectivity.png}{(\zoomxBaseball, \zoomyBaseball)}{south east} &
        \zoomDepth{label}{images/ablations/baseball/no\_bias\_depth.png}{(\zoomxBaseball, \zoomyBaseball)}{south east} &
        \zoomDepth{label}{images/ablations/baseball/+dd\_depth.png}{(\zoomxBaseball, \zoomyBaseball)}{south east} &
        \zoomDepth{label}{images/ablations/baseball/+h1\_depth.png}{(\zoomxBaseball, \zoomyBaseball)}{south east} &
        \zoomDepth{label}{images/ablations/baseball/+h1\_+h2\_depth.png}{(\zoomxBaseball, \zoomyBaseball)}{south east} &
        \zoomDepth{label}{images/ablations/baseball/+h1\_+h2\_+dd\_depth.png}{(\zoomxBaseball, \zoomyBaseball)}{south east} \\
    \end{tabular}
    
    \vspace{-3.5mm}
    
    \caption{\textbf{Ablations}. In the first column, the even rows show the reflectivity map, which is computed as input amplitude multiplied with the square of input depth (light falloff), and can be understood as the expected of Gaussian reflectivity at the corresponding surface. For visualization, the map is clipped to the range [0, 1], where overexposed areas only indicate high target reflectivity. No bias led to arbitrary number of density peaks even for opaque surfaces, thus inaccurate depth from 
    \maineq{5}.
    Adding DD overly stack large Gaussians that led to overly smoothed mean depth. Occupancy bias improves the placement of Gaussians but still struggles in low-reflectivity areas (e.g., bottom left of the cart, and the pillow). Initializing Gaussians with low reflectivity mitigates this issue. Reintroducing the DD loss again after applying the two heuristics still led to oversmooth depth due to Gaussians' spatial extent.}
    \label{fig:qualitative-ablations-combined}
\end{figure*}
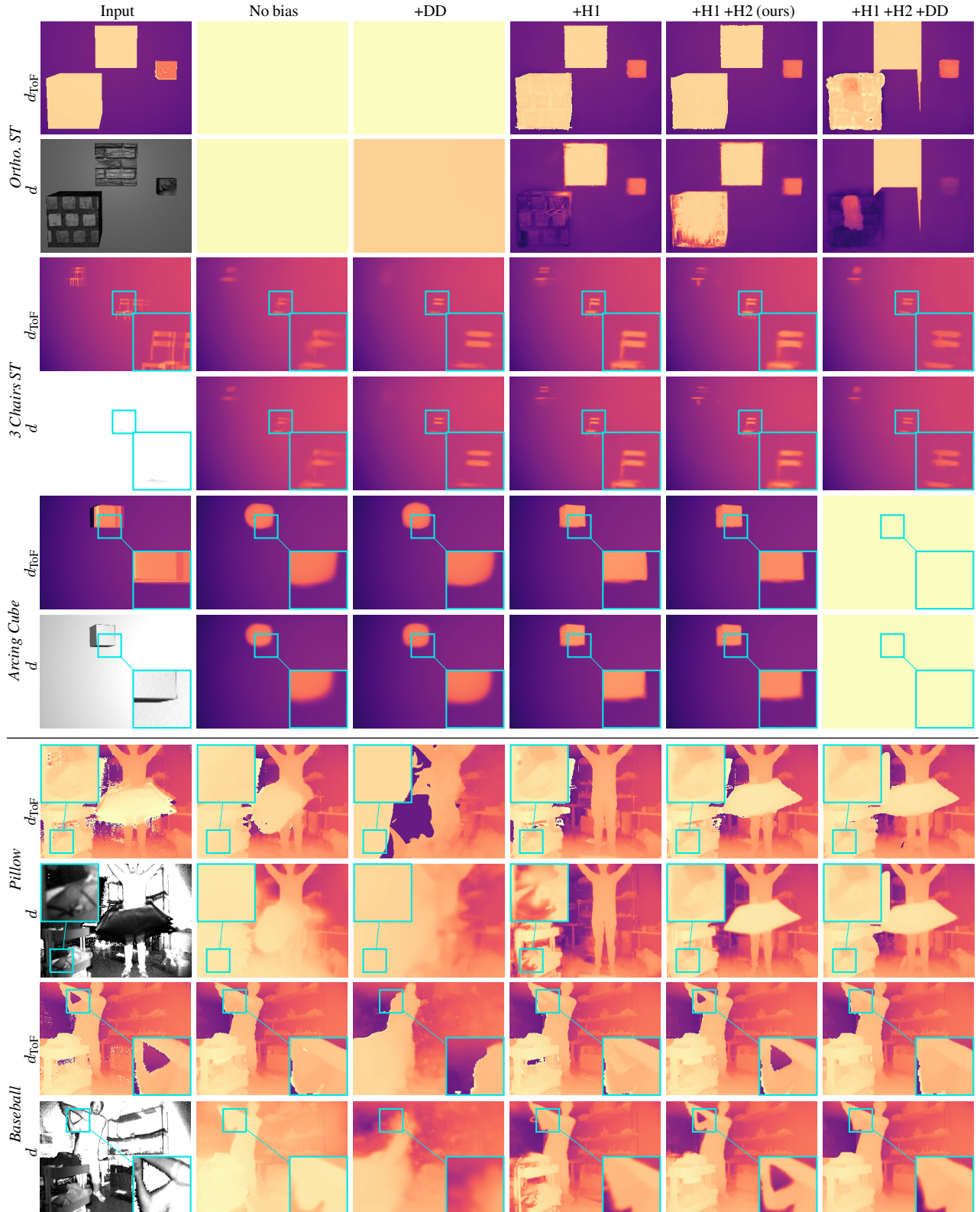

We show quantitative (Tab. \ref{quantitative-synthetic-table-ablations}) ablations on \ftorf synthetic scenes and also qualitative ablation results on \ftorf real-world and synthetic scenes (Fig. \ref{fig:qualitative-ablations-combined}). ``No bias'' refers to no consideration of either of our heuristics. ``DD'' means adding the depth distortion loss \cite{huang20242d}, and ``H1'' and ``H2'' means our control of the optimization using our occupancy bias (H1) and low initial reflectivity bias (H2). For result variants, `$d$' refers to the mean depth from \maineq{5}, and `$d_\text{ToF}$' refers to the depth derived from the reconstructed quads using \maineq{1}.

\begin{table*}[t]
    \caption{\textbf{\label{quantitative-synthetic-table-ablations} Ablation depth error on the \ftorf synthetic dataset}. Each number is a depth MSE$\times 100$.
    Bold marks the smallest result for each of $d$ and $d_\text{ToF}$. On average, our approach is reliably more accurate without any catastrophic failures, though scene specific variations exist. \SeqSlidingCube is too simple to induce large differences. \SeqZMotionShort is the only sequence with large low reflectivity areas, and we see the advantage of our low initial reflectivity bias (H2) in the scene's reconstructed $d$. Using both heuristics H1, H2 plus the depth distortion (DD) loss theoretically should reduce the gap between $d$ and $d_\text{ToF}$, but in practice, because of the spatial extent of the Gaussians, this usually led to oversmooth depth, worse thin structures, and is sometimes unstable because of the existence of large Gaussians representing large homogeneous textureless areas (\cref{fig:qualitative-ablations-combined}).}
    \centering
    \small
    \begin{tabular}{l | l | r | r | r | r | r | r | r | r}
        \toprule
        & & Mean & {\SeqSlidingCube} & {\emph{Occ. Cube}} & {\SeqAxialShort} & {\SeqThreeCubesShort} & {\SeqThreeChairsShort} & {\SeqArcingCube} & {\SeqZMotionShort} \\
        \midrule
        \multirow{5}*{\raisebox{0.0in}{\rotatebox[origin=c]{90}{\(d_{\text{{ToF}}}\)\xspace}}}
        & No bias & 144.134 & {0.013} & {13.227} & {0.286} & {1.152} & {0.885} & {5.884} & {987.492} \\
        & +DD & 146.189 & {0.119} & {11.422} & {5.413} & {7.289} & {1.224} & {13.415} & {984.440} \\
        & +H1 & 3.641 & {0.017} & {\textbf{0.024}} & {0.206} & {0.916} & {\textbf{0.605}} & {\textbf{1.113}} & {22.610} \\
        & +H1 +H2 (ours) & {{\textbf{3.360}}} & {\textbf{0.008}} & {0.073} & {\textbf{0.154}} & {\textbf{0.294}} & {0.650} & {1.526} & {\textbf{20.818}} \\
        & +H1 +H2 +DD & 255.717 & {0.012} & {148.578} & {0.422} & {1.010} & {0.895} & {1537.952} & {101.148} \\        
        \arrayrulecolor{black!100}\midrule
        \midrule
        \multirow{5}*{\raisebox{0.0in}{\rotatebox[origin=c]{90}{\(d\)\xspace}}}
        & No bias & 365.686 & {0.049} & {15.604} & {55.834} & {24.892} & {1.482} & {4.456} & {2457.487} \\
        & +DD & 274.316 & {0.172} & {11.675} & {5.922} & {7.805} & {1.421} & {13.067} & {1880.148} \\
        & +H1 & 44.520 & {0.058} & {\textbf{0.310}} & {\textbf{0.726}} & {1.610} & {\textbf{1.011}} & {\textbf{1.025}} & {306.898} \\
        & +H1 +H2 (ours) & {\textbf{5.824}} & {0.033} & {0.348} & {0.730} & {\textbf{1.059}} & {1.072} & {1.479} & {\textbf{36.048}} \\
        & +H1 +H2 +DD & 310.564 & {\textbf{0.026}} & {147.856} & {0.787} & {1.362} & {1.106} & {1537.840} & {484.997} \\
        \bottomrule
    \end{tabular}
\end{table*}

\renewcommand{\theequation}{S\arabic{equation}}

\section{ToF Gradient Computation Details}
We briefly explain the opacity gradient computation for our ToF image formation model, which is less trivial compared to other gradients computed via simple chain rules. Let \( L \) denote the loss function, and \( \alpha_k = o_k \mathcal{G}^{2D}_k(\mathbf{x}) \) the opacity term of the \(k\)-th Gaussian. Using the chain rule:
\begin{align}
    \frac{\partial L}{\partial \alpha_k} = \frac{\partial L}{\partial \mathbf{c}(\mathbf{x})} \frac{\partial \mathbf{c}(\mathbf{x})}{\partial \alpha_k},
\end{align}
the term \( \frac{\partial \mathbf{c}(\mathbf{x})}{\partial \alpha_k} \) is computed recursively in the original 3D Gaussian Splatting (3DGS) method \cite{kerbl20233d}:
\begin{align}
    \frac{\partial \mathbf{c}(\mathbf{x})}{\partial \alpha_k} &= T_k (\mathbf{c}_k - \text{acc}_k) \text{,} \\
    \text{acc}_k &=
    \begin{cases}
        \alpha_{k+1} \mathbf{c}_{k+1} + (1 - \alpha_{k+1}) \text{acc}_{k+1}, & k < N , \\
        0, & k = N,
    \end{cases}  \nonumber
\end{align}
where \( T_k \) is the transmittance, \( \mathbf{c}_k \) is the Gaussian’s color, and \( \text{acc}_k \) aggregates contributions of later Gaussians.

In our model, this computation extends to C-ToF signals like phasor or quad pixels (\( \mathbf{p}(\mathbf{x}) \), \( \mathbf{q}(\mathbf{x}) \)) as described in 
\maineq{4}.
We take \( \mathbf{q}(\mathbf{x}) \) as an example, the corresponding recursion becomes to:
\begin{align}
    \frac{\partial \mathbf{q}(\mathbf{x})}{\partial \alpha_k} &= T_k^2 (\mathbf{q}_k + 2 (\alpha_k - 1) \text{acc}_k^q) \text{,} \\
    \text{acc}_k^q &=
    \begin{cases}
        \alpha_{k+1} \mathbf{q}_{k+1} + (1 - \alpha_{k+1})^2 \text{acc}_{k+1}^q, & k < N, \\
        0, & k = N.
    \end{cases}  \nonumber
\end{align}
Here, \( \mathbf{q}_k \) represents the quad contribution of the \(k\)-th Gaussian, and \( \text{acc}_k^q \) accumulates later quad contributions.

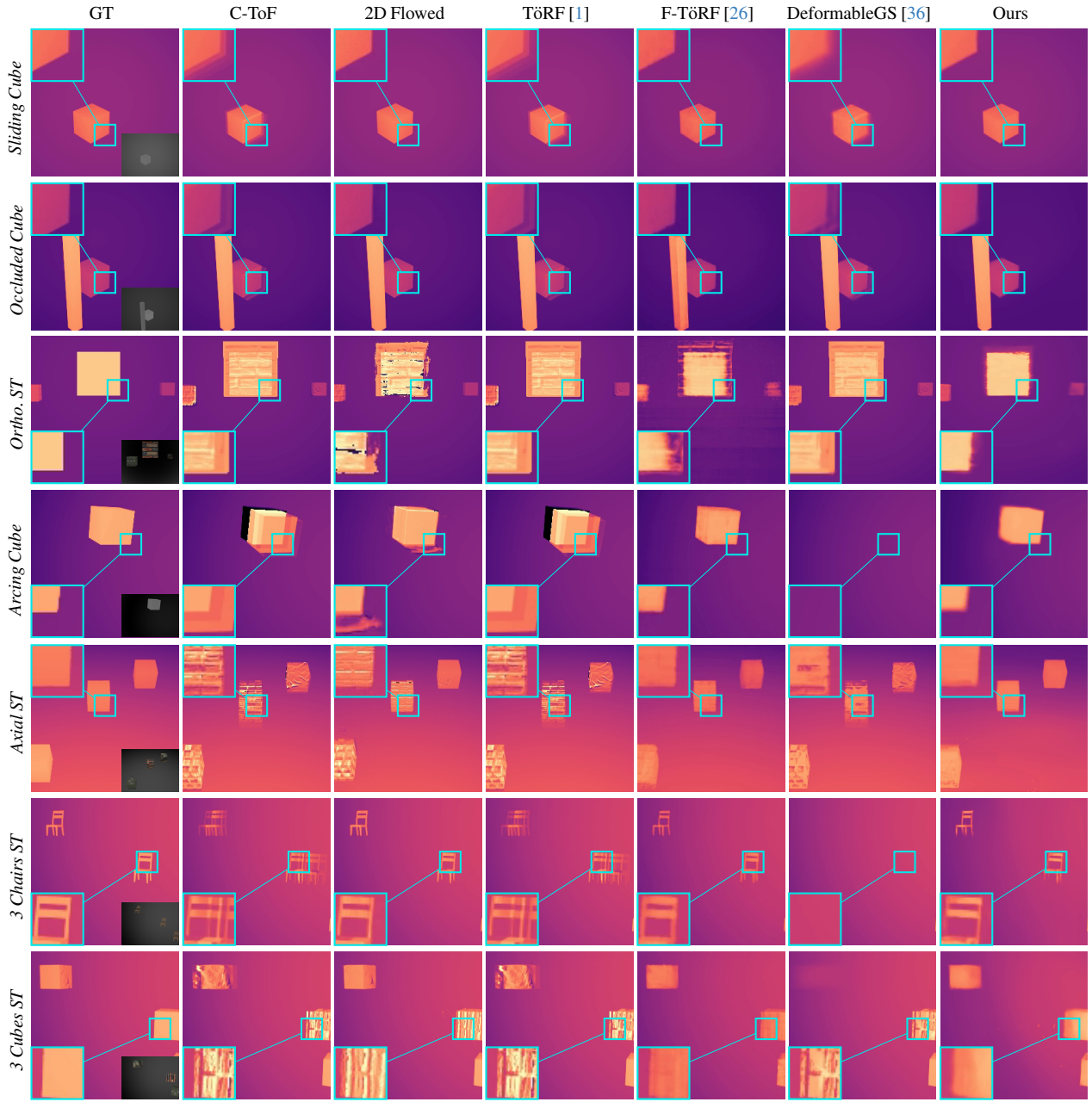
\begin{figure*}
    \centering
    \footnotesize
    \vspace{-2mm}

    \newcommand{\imgw}{0.13\linewidth}
    \newcommand{\imgwtiny}{0.050\linewidth} 

    \newcommand{\zoomxSlidingCube}{0.0} 
    \newcommand{\zoomySlidingCube}{0.5} 

    \newcommand{\zoomxOcclusion}{0.0} 
    \newcommand{\zoomyOcclusion}{0.4} 

    \newcommand{\zoomxAcuteZSpeedTest}{-0.2} 
    \newcommand{\zoomyAcuteZSpeedTest}{-0.3} 

    \newcommand{\zoomxArcingCube}{-0.4} 
    \newcommand{\zoomyArcingCube}{-0.3} 

    \newcommand{\zoomxZMotionSpeedTest}{0.0} 
    \newcommand{\zoomyZMotionSpeedTest}{-0.2} 

    \newcommand{\zoomxSpeedTestChair}{-0.65} 
    \newcommand{\zoomySpeedTestChair}{-0.15} 

    \newcommand{\zoomxSpeedTestTexture}{-0.85} 
    \newcommand{\zoomySpeedTestTexture}{0.05} 

    \newcommand{\imgDepth}[2][0px 0px 0px 0px]{\includegraphics[draft=\isdraft,width=\imgw,trim=#1,clip,viewport=50 0 290 240]{#2}}
    \newcommand{\zoomDepth}[4]{
        \trimbox{0cm 0cm 0.15cm 0.0cm}{
        \begin{tikzpicture}[
    		image/.style={inner sep=0pt, outer sep=0pt},
    		collabel/.style={above=9pt, anchor=north, inner ysep=0pt, align=center}, 
    		rowlabel/.style={left=9pt, rotate=90, anchor=north, inner ysep=0pt, scale=0.8, align=center},
    		subcaption/.style={inner xsep=0.75mm, inner ysep=0.75mm, below right},
    		arrow/.style={-{Latex[length=2.5mm,width=4mm]}, line width=2mm},
    		spy using outlines={rectangle, size=0.8cm, magnification=2.5, connect spies, ultra thick, every spy on node/.append style={thick}},
    		style1/.style={cyan!90!black,thick},
    		style2/.style={orange!90!black},
    		style3/.style={blue!90!black},
    		style4/.style={green!90!black},
    		style5/.style={white},
    		style6/.style={black},
        ]
        
        \node [image] (#1) {\imgDepth[0px 0px 0px 0px]{#2}};
        \spy[style1] on ($(#1.center)-#3$) in node (crop-#1) [anchor=#4] at (#1.#4);
        
        \end{tikzpicture}
        }
    }

    \newcommand{\imgRGBTiny}[2][0px 0px 0px 0px]{\includegraphics[draft=\isdraft,width=\imgwtiny,trim=#1,clip]{#2}}
    \newcommand{\zoomDepthTinyRGB}[5]{
        \trimbox{0cm 0cm 0.15cm 0.0cm}{
        \begin{tikzpicture}[
    		image/.style={inner sep=0pt, outer sep=0pt},
                image2/.style={inner sep=0pt, outer sep=0pt},
    		collabel/.style={above=9pt, anchor=north, inner ysep=0pt, align=center}, 
    		rowlabel/.style={left=9pt, rotate=90, anchor=north, inner ysep=0pt, scale=0.8, align=center},
    		subcaption/.style={inner xsep=0.75mm, inner ysep=0.75mm, below right},
    		arrow/.style={-{Latex[length=2.5mm,width=4mm]}, line width=2mm},
    		spy using outlines={rectangle, size=0.8cm, magnification=2.5, connect spies, ultra thick, every spy on node/.append style={thick}},
    		style1/.style={cyan!90!black,thick},
    		style2/.style={orange!90!black},
    		style3/.style={blue!90!black},
    		style4/.style={green!90!black},
    		style5/.style={white},
    		style6/.style={black},
        ]
        
        \node [image] (#1) {\imgDepth[0px 0px 0px 0px]{#2}};
        \spy[style1] on ($(#1.center)-#3$) in node (crop-#1) [anchor=#4] at (#1.#4);

        \node [image2] at (-19.8pt,22.8pt) {\imgRGBTiny[0px 0px 0px 0px]{#5}};
        
        \end{tikzpicture}
        }
    }

    \newcommand{\zoomDepthTinyRGBbottom}[5]{
        \trimbox{0cm 0cm 0.15cm 0.0cm}{
        \begin{tikzpicture}[
    		image/.style={inner sep=0pt, outer sep=0pt},
                image2/.style={inner sep=0pt, outer sep=0pt},
    		collabel/.style={above=9pt, anchor=north, inner ysep=0pt, align=center}, 
    		rowlabel/.style={left=9pt, rotate=90, anchor=north, inner ysep=0pt, scale=0.8, align=center},
    		subcaption/.style={inner xsep=0.75mm, inner ysep=0.75mm, below right},
    		arrow/.style={-{Latex[length=2.5mm,width=4mm]}, line width=2mm},
    		spy using outlines={rectangle, size=0.8cm, magnification=2.5, connect spies, ultra thick, every spy on node/.append style={thick}},
    		style1/.style={cyan!90!black,thick},
    		style2/.style={orange!90!black},
    		style3/.style={blue!90!black},
    		style4/.style={green!90!black},
    		style5/.style={white},
    		style6/.style={black},
        ]
        
        \node [image] (#1) {\imgDepth[0px 0px 0px 0px]{#2}};
        \spy[style1] on ($(#1.center)-#3$) in node (crop-#1) [anchor=#4] at (#1.#4);

        \node [image2] at (20pt,-22.9pt) {\imgRGBTiny[0px 0px 0px 0px]{#5}};
        
        \end{tikzpicture}
        }
    }

    \setlength{\tabcolsep}{0.5pt}
    \begin{tabular}{cccccccc}
    &
    GT &
    C-ToF &
    2D Flowed &
    \torf \cite{attal2021torf} &
    \ftorf \cite{okunev2024flowed} &
    DeformableGS \cite{yang2024deformable} &
    Ours
    \\[0.2em]
    
    {\raisebox{0.40in}{\rotatebox[origin=c]{90}{\SeqSlidingCube}}} &
    \zoomDepthTinyRGBbottom{label0}{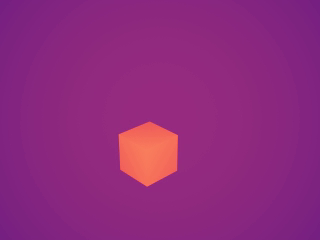}{(\zoomxSlidingCube,\zoomySlidingCube)}{north west}
    {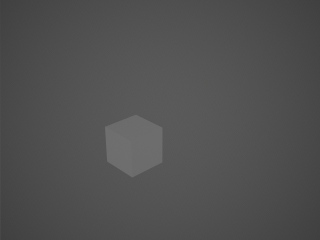}
    &
    \zoomDepth{label1}{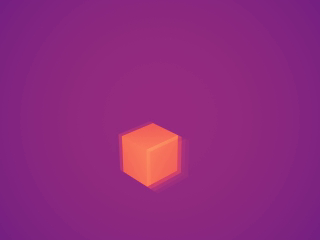}{(\zoomxSlidingCube,\zoomySlidingCube)}{north west} &
    \zoomDepth{label2}{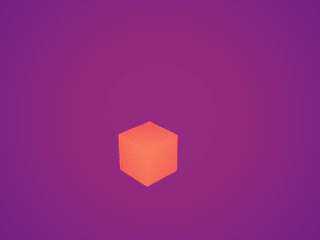}{(\zoomxSlidingCube,\zoomySlidingCube)}{north west} &
    \zoomDepth{label3}{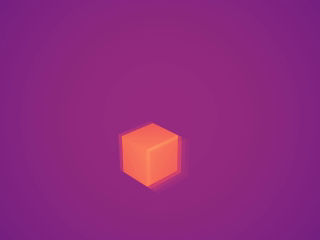}{(\zoomxSlidingCube,\zoomySlidingCube)}{north west} &
    \zoomDepth{label4}{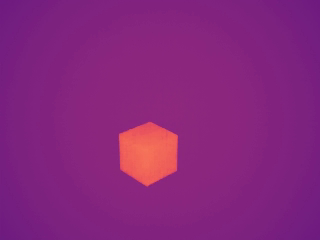}{(\zoomxSlidingCube,\zoomySlidingCube)}{north west} &
    \zoomDepth{label5}{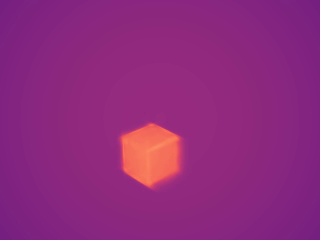}{(\zoomxSlidingCube,\zoomySlidingCube)}{north west} &
    \zoomDepth{label5}{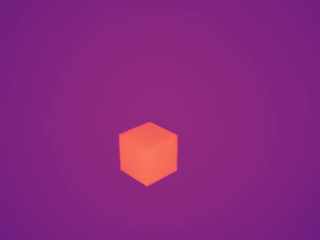}{(\zoomxSlidingCube,\zoomySlidingCube)}{north west} \\
    [0.0em]
    
    {\raisebox{0.4in}{\rotatebox[origin=c]{90}{\SeqOccludedCube}}} &
    \zoomDepthTinyRGBbottom{label0}{images/qualitative_comparison/ftorf_synthetic/occlusion/gt.png}{(\zoomxOcclusion,\zoomyOcclusion)}{north west} 
    {images/rgb_gt/occlusion.png}
    &
    \zoomDepth{label1}{images/qualitative_comparison/ftorf_synthetic/occlusion/ctof.png}{(\zoomxOcclusion,\zoomyOcclusion)}{north west} &
    \zoomDepth{label2}{images/qualitative_comparison/ftorf_synthetic/occlusion/warped.png}{(\zoomxOcclusion,\zoomyOcclusion)}{north west} &
    \zoomDepth{label3}{images/qualitative_comparison/ftorf_synthetic/occlusion/torf.png}{(\zoomxOcclusion,\zoomyOcclusion)}{north west} &
    \zoomDepth{label4}{images/qualitative_comparison/ftorf_synthetic/occlusion/ftorf.png}{(\zoomxOcclusion,\zoomyOcclusion)}{north west} &
    \zoomDepth{label5}{images/qualitative_comparison/ftorf_synthetic/occlusion/deformablegs_depth.png}{(\zoomxOcclusion,\zoomyOcclusion)}{north west} &
    \zoomDepth{label5}{images/qualitative_comparison/ftorf_synthetic/occlusion/gtorf_depth.png}{(\zoomxOcclusion,\zoomyOcclusion)}{north west} \\
    [0.0em]

    {\raisebox{0.4in}{\rotatebox[origin=c]{90}{\SeqZMotionShort}}} &
    \zoomDepthTinyRGBbottom{label0}{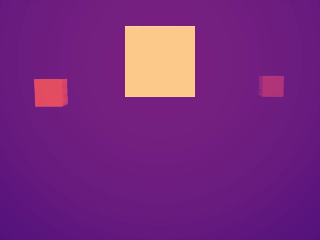}{(\zoomxAcuteZSpeedTest,\zoomyAcuteZSpeedTest)}{south west}
    {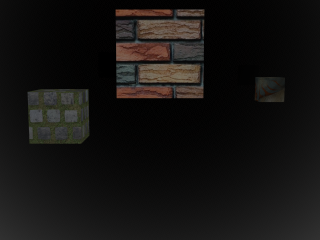}
    &
    \zoomDepth{label1}{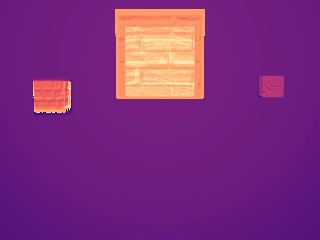}{(\zoomxAcuteZSpeedTest,\zoomyAcuteZSpeedTest)}{south west} &
    \zoomDepth{label2}{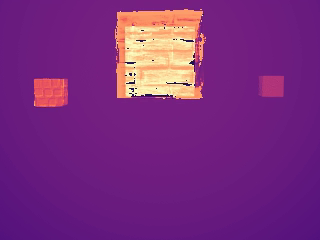}{(\zoomxAcuteZSpeedTest,\zoomyAcuteZSpeedTest)}{south west} &
    \zoomDepth{label3}{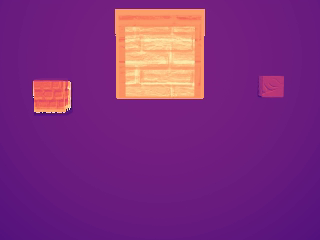}{(\zoomxAcuteZSpeedTest,\zoomyAcuteZSpeedTest)}{south west} &
    \zoomDepth{label4}{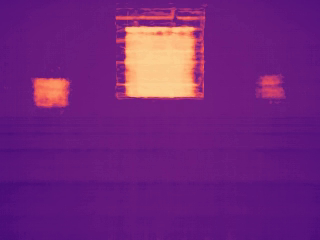}{(\zoomxAcuteZSpeedTest,\zoomyAcuteZSpeedTest)}{south west} &
    \zoomDepth{label5}{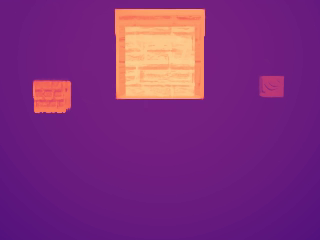}{(\zoomxAcuteZSpeedTest,\zoomyAcuteZSpeedTest)}{south west} &
    \zoomDepth{label5}{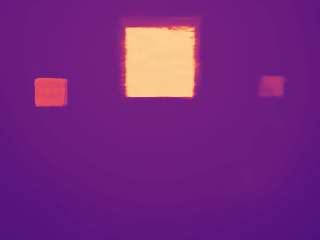}{(\zoomxAcuteZSpeedTest,\zoomyAcuteZSpeedTest)}{south west} \\
    [0.0em]

    {\raisebox{0.4in}{\rotatebox[origin=c]{90}{\SeqArcingCube}}} &
    \zoomDepthTinyRGBbottom{label0}{images/qualitative_comparison/ftorf_synthetic/arcing_cube/gt.png}{(\zoomxArcingCube,\zoomyArcingCube)}{south west}
    {images/rgb_gt/arcing_cube.png}
    &
    \zoomDepth{label1}{images/qualitative_comparison/ftorf_synthetic/arcing_cube/ctof.png}{(\zoomxArcingCube,\zoomyArcingCube)}{south west} &
    \zoomDepth{label2}{images/qualitative_comparison/ftorf_synthetic/arcing_cube/warped.png}{(\zoomxArcingCube,\zoomyArcingCube)}{south west} &
    \zoomDepth{label3}{images/qualitative_comparison/ftorf_synthetic/arcing_cube/torf.png}{(\zoomxArcingCube,\zoomyArcingCube)}{south west} &
    \zoomDepth{label4}{images/qualitative_comparison/ftorf_synthetic/arcing_cube/ftorf.png}{(\zoomxArcingCube,\zoomyArcingCube)}{south west} &
    \zoomDepth{label5}{images/qualitative_comparison/ftorf_synthetic/arcing_cube/deformablegs_depth.png}{(\zoomxArcingCube,\zoomyArcingCube)}{south west} &
    \zoomDepth{label5}{images/qualitative_comparison/ftorf_synthetic/arcing_cube/gtorf_depth.png}{(\zoomxArcingCube,\zoomyArcingCube)}{south west} \\
    [0.0em]

    {\raisebox{0.4in}{\rotatebox[origin=c]{90}{\SeqAxialShort}}} &
    \zoomDepthTinyRGBbottom{label1}{images/qualitative_comparison/ftorf_synthetic/z_motion_speed_test/gt.png}{(\zoomxZMotionSpeedTest,\zoomyZMotionSpeedTest)}{north west}
    {images/rgb_gt/z_motion_speed_test}
    &
    \zoomDepth{label1}{images/qualitative_comparison/ftorf_synthetic/z_motion_speed_test/ctof.png}{(\zoomxZMotionSpeedTest,\zoomyZMotionSpeedTest)}{north west} &
    \zoomDepth{label2}{images/qualitative_comparison/ftorf_synthetic/z_motion_speed_test/warped.png}{(\zoomxZMotionSpeedTest,\zoomyZMotionSpeedTest)}{north west} &
    \zoomDepth{label3}{images/qualitative_comparison/ftorf_synthetic/z_motion_speed_test/torf.png}{(\zoomxZMotionSpeedTest,\zoomyZMotionSpeedTest)}{north west} &
    \zoomDepth{label4}{images/qualitative_comparison/ftorf_synthetic/z_motion_speed_test/ftorf.png}{(\zoomxZMotionSpeedTest,\zoomyZMotionSpeedTest)}{north west} &
    \zoomDepth{label5}{images/qualitative_comparison/ftorf_synthetic/z_motion_speed_test/deformablegs_depth.png}{(\zoomxZMotionSpeedTest,\zoomyZMotionSpeedTest)}{north west} &
    \zoomDepth{label5}{images/qualitative_comparison/ftorf_synthetic/z_motion_speed_test/gtorf_depth.png}{(\zoomxZMotionSpeedTest,\zoomyZMotionSpeedTest)}{north west} \\
    [0.0em]

    {\raisebox{0.4in}{\rotatebox[origin=c]{90}{\SeqThreeChairsShort}}} &
    \zoomDepthTinyRGBbottom{label0}{images/qualitative_comparison/ftorf_synthetic/speed_test_chair/gt.png}{(\zoomxSpeedTestChair,\zoomySpeedTestChair)}{south west}
    {images/rgb_gt/speed_test_chair.png}
    &
    \zoomDepth{label1}{images/qualitative_comparison/ftorf_synthetic/speed_test_chair/ctof.png}{(\zoomxSpeedTestChair,\zoomySpeedTestChair)}{south west} &
    \zoomDepth{label2}{images/qualitative_comparison/ftorf_synthetic/speed_test_chair/warped.png}{(\zoomxSpeedTestChair,\zoomySpeedTestChair)}{south west} &
    \zoomDepth{label3}{images/qualitative_comparison/ftorf_synthetic/speed_test_chair/torf.png}{(\zoomxSpeedTestChair,\zoomySpeedTestChair)}{south west} &
    \zoomDepth{label4}{images/qualitative_comparison/ftorf_synthetic/speed_test_chair/ftorf.png}{(\zoomxSpeedTestChair,\zoomySpeedTestChair)}{south west} &
    \zoomDepth{label5}{images/qualitative_comparison/ftorf_synthetic/speed_test_chair/deformablegs_depth.png}{(\zoomxSpeedTestChair,\zoomySpeedTestChair)}{south west} &
    \zoomDepth{label5}{images/qualitative_comparison/ftorf_synthetic/speed_test_chair/gtorf_depth.png}{(\zoomxSpeedTestChair,\zoomySpeedTestChair)}{south west} \\
    [0.0em]
    
    {\raisebox{0.4in}{\rotatebox[origin=c]{90}{\SeqThreeCubesShort}}} &
    \zoomDepthTinyRGBbottom{label0}{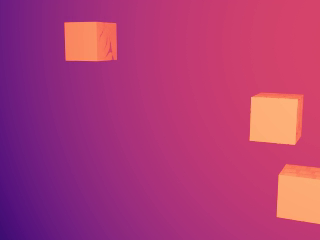}{(\zoomxSpeedTestTexture,\zoomySpeedTestTexture)}{south west} 
    {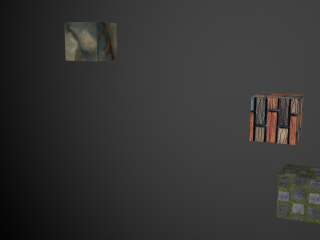}
    &
    \zoomDepth{label1}{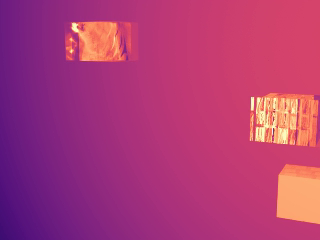}{(\zoomxSpeedTestTexture,\zoomySpeedTestTexture)}{south west} &
    \zoomDepth{label2}{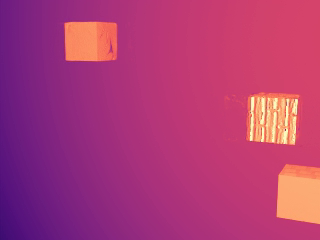}{(\zoomxSpeedTestTexture,\zoomySpeedTestTexture)}{south west} &
    \zoomDepth{label3}{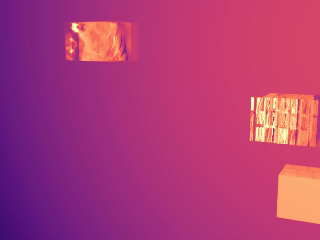}{(\zoomxSpeedTestTexture,\zoomySpeedTestTexture)}{south west} &
    \zoomDepth{label4}{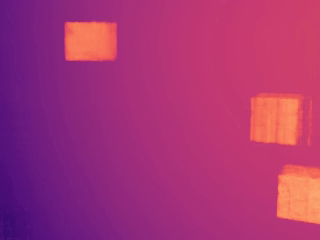}{(\zoomxSpeedTestTexture,\zoomySpeedTestTexture)}{south west} &
    \zoomDepth{label5}{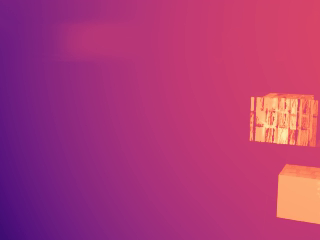}{(\zoomxSpeedTestTexture,\zoomySpeedTestTexture)}{south west} &
    \zoomDepth{label5}{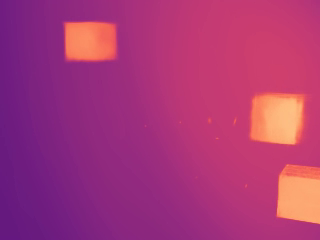}{(\zoomxSpeedTestTexture,\zoomySpeedTestTexture)}{south west} \\
    
    \end{tabular}
    \vspace{-2mm}
    \caption{\textbf{Results on all \ftorf synthetic scenes}. DeformableGS is given C-ToF-derived depth as an additional input. \textit{Inset on ground truth: corresponding RGB image. Full RGB images can be found in \cref{fig:synthetic_rgb_legend}.}}
    \vspace{-1em}
    \label{fig:real-synthetic-baselines-supp}
\end{figure*}
\begin{figure*}
    \centering
    \footnotesize

    \newcommand{\imgw}{0.15\linewidth}

    \newcommand{\zoomxBaseball}{-0.4} 
    \newcommand{\zoomyBaseball}{-0.45} 

    \newcommand{\zoomxPillow}{-1.05} 
    \newcommand{\zoomyPillow}{0.4} 

    \newcommand{\zoomxJacks}{1.0} 
    \newcommand{\zoomyJacks}{-0.95} 

    \newcommand{\zoomxTarget}{-0.1} 
    \newcommand{\zoomyTarget}{-0.1} 

    \newcommand{\zoomxFan}{-0.25} 
    \newcommand{\zoomyFan}{0.25} 

    \newcommand{\imgDepth}[2][0px 0px 0px 0px]{\includegraphics[draft=\isdraft,width=\imgw,trim=#1,clip,viewport=50 0 290 240]{#2}}
    \newcommand{\zoomDepth}[4]{
        \trimbox{0cm 0cm 0.15cm 0.0cm}{
        \begin{tikzpicture}[
    		image/.style={inner sep=0pt, outer sep=0pt},
    		collabel/.style={above=9pt, anchor=north, inner ysep=0pt, align=center}, 
    		rowlabel/.style={left=9pt, rotate=90, anchor=north, inner ysep=0pt, scale=0.8, align=center},
    		subcaption/.style={inner xsep=0.75mm, inner ysep=0.75mm, below right},
    		arrow/.style={-{Latex[length=2.5mm,width=4mm]}, line width=2mm},
    		spy using outlines={rectangle, size=1.25cm, magnification=2.5, connect spies, ultra thick, every spy on node/.append style={thick}},
    		style1/.style={cyan!90!black,thick},
    		style2/.style={orange!90!black},
    		style3/.style={blue!90!black},
    		style4/.style={green!90!black},
    		style5/.style={white},
    		style6/.style={black},
        ]
        
        \node [image] (#1) {\imgDepth[0px 0px 0px 0px]{#2}};
        \spy[style1] on ($(#1.center)-#3$) in node (crop-#1) [anchor=#4] at (#1.#4);
        
        \end{tikzpicture}
        }
    }

    \vspace{1mm}
    \setlength{\tabcolsep}{0.5pt}
    \begin{tabular}{ccccccc}
    &
    C-ToF &
    2D Flowed &
    \torf \cite{attal2021torf} &
    \ftorf \cite{okunev2024flowed} &
    DeformableGS \cite{yang2024deformable} &
    Ours
    \\[0.2em]
    
    {\raisebox{0.5in}{\rotatebox[origin=c]{90}{\SeqBaseball}}} &
    \zoomDepth{label1}{images/qualitative_comparison/ftorf_real/baseball/ctof.png}{(\zoomxBaseball,\zoomyBaseball)}{south east} &
    \zoomDepth{label2}{images/qualitative_comparison/ftorf_real/baseball/warped.png}{(\zoomxBaseball,\zoomyBaseball)}{south east} &
    \zoomDepth{label3}{images/qualitative_comparison/ftorf_real/baseball/torf.png}{(\zoomxBaseball,\zoomyBaseball)}{south east} &
    \zoomDepth{label4}{images/qualitative_comparison/ftorf_real/baseball/ftorf.png}{(\zoomxBaseball,\zoomyBaseball)}{south east} &
    \zoomDepth{label5}{images/qualitative_comparison/ftorf_real/baseball/deformablegs_depth.png}{(\zoomxBaseball,\zoomyBaseball)}{south east} &
    \zoomDepth{label5}{images/qualitative_comparison/ftorf_real/baseball/gtorf_depth.png}{(\zoomxBaseball,\zoomyBaseball)}{south east} \\

    {\raisebox{0.5in}{\rotatebox[origin=c]{90}{\SeqPillow}}} &
    \zoomDepth{label1}{images/qualitative_comparison/ftorf_real/pillow/ctof.png}{(\zoomxPillow,\zoomyPillow)}{south west} &
    \zoomDepth{label2}{images/qualitative_comparison/ftorf_real/pillow/warped.png}{(\zoomxPillow,\zoomyPillow)}{south west} &
    \zoomDepth{label3}{images/qualitative_comparison/ftorf_real/pillow/torf.png}{(\zoomxPillow,\zoomyPillow)}{south west} &
    \zoomDepth{label4}{images/qualitative_comparison/ftorf_real/pillow/ftorf.png}{(\zoomxPillow,\zoomyPillow)}{south west} &
    \zoomDepth{label5}{images/qualitative_comparison/ftorf_real/pillow/deformablegs_depth.png}{(\zoomxPillow,\zoomyPillow)}{south west} &
    \zoomDepth{label5}{images/qualitative_comparison/ftorf_real/pillow/gtorf_depth.png}{(\zoomxPillow,\zoomyPillow)}{south west} \\

    {\raisebox{0.5in}{\rotatebox[origin=c]{90}{\SeqJumpingJacks}}} &
    \zoomDepth{label1}{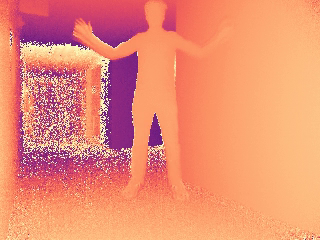}{(\zoomxJacks,\zoomyJacks)}{south east} &
    \zoomDepth{label2}{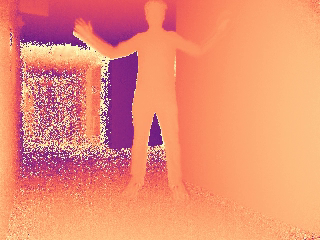}{(\zoomxJacks,\zoomyJacks)}{south east} &
    \zoomDepth{label3}{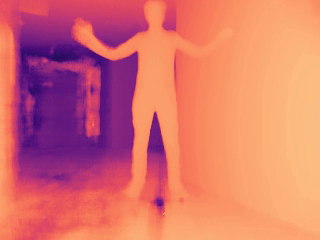}{(\zoomxJacks,\zoomyJacks)}{south east} &
    \zoomDepth{label4}{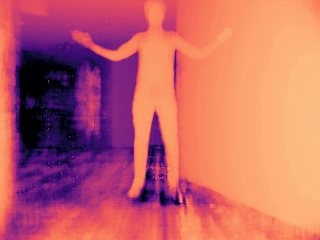}{(\zoomxJacks,\zoomyJacks)}{south east} &
    \zoomDepth{label5}{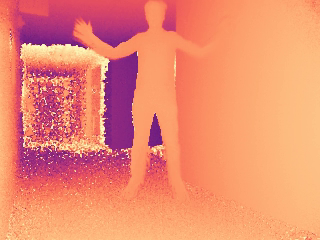}{(\zoomxJacks,\zoomyJacks)}{south east} &
    \zoomDepth{label5}{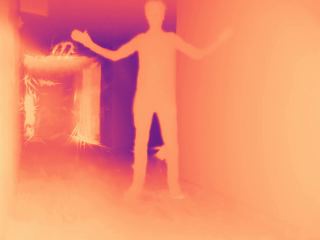}{(\zoomxJacks,\zoomyJacks)}{south east} \\

    {\raisebox{0.5in}{\rotatebox[origin=c]{90}{\SeqTarget}}} &
    \zoomDepth{label1}{images/qualitative_comparison/ftorf_real/target/ctof.png}{(\zoomxTarget,\zoomyTarget)}{north west} &
    \zoomDepth{label2}{images/qualitative_comparison/ftorf_real/target/warped.png}{(\zoomxTarget,\zoomyTarget)}{north west} &
    \zoomDepth{label3}{images/qualitative_comparison/ftorf_real/target/torf.png}{(\zoomxTarget,\zoomyTarget)}{north west} &
    \zoomDepth{label4}{images/qualitative_comparison/ftorf_real/target/ftorf.png}{(\zoomxTarget,\zoomyTarget)}{north west} &
    \zoomDepth{label5}{images/qualitative_comparison/ftorf_real/target/deformablegs_depth.png}{(\zoomxTarget,\zoomyTarget)}{north west} &
    \zoomDepth{label5}{images/qualitative_comparison/ftorf_real/target/0028.png}{(\zoomxTarget,\zoomyTarget)}{north west}  \\
    
    {\raisebox{0.5in}{\rotatebox[origin=c]{90}{\SeqFan}}} &
    \zoomDepth{label1}{images/qualitative_comparison/ftorf_real/fan/ctof.png}{(\zoomxFan,\zoomyFan)}{south west} &
    \zoomDepth{label2}{images/qualitative_comparison/ftorf_real/fan/warped.png}{(\zoomxFan,\zoomyFan)}{south west} &
    \zoomDepth{label3}{images/qualitative_comparison/ftorf_real/fan/torf.png}{(\zoomxFan,\zoomyFan)}{south west} &
    \zoomDepth{label4}{images/qualitative_comparison/ftorf_real/fan/ftorf.png}{(\zoomxFan,\zoomyFan)}{south west} &
    \zoomDepth{label5}{images/qualitative_comparison/ftorf_real/fan/deformablegs_depth.png}{(\zoomxFan,\zoomyFan)}{south west} &
    \zoomDepth{label5}{images/qualitative_comparison/ftorf_real/fan/gtorf_depth.png}{(\zoomxFan,\zoomyFan)}{south west} \\

    \end{tabular}
    \vspace{-2mm}
    \caption{\textbf{Results on all \ftorf real-world scenes}. DeformableGS is given C-ToF-derived depth as an additional input.}
    \label{fig:real-qualitative-baselines-supp}
\end{figure*}
\begin{figure*}
    \centering
    \footnotesize

    \newcommand{\imgw}{0.15\linewidth}

    \newcommand{\zoomxCupboard}{0.2} 
    \newcommand{\zoomyCupboard}{-0.1} 

    \newcommand{\zoomxDeskbox}{-0.4} 
    \newcommand{\zoomyDeskbox}{-0.5} 

    \newcommand{\zoomxCopier}{-0.62} 
    \newcommand{\zoomyCopier}{-0.75} 

    \newcommand{\zoomxStudybook}{0.1} 
    \newcommand{\zoomyStudybook}{-0.2} 

    \newcommand{\imgDepth}[2][0px 0px 0px 0px]{\includegraphics[draft=\isdraft,width=\imgw,trim=#1,clip,viewport=50 0 290 240]{#2}}
    \newcommand{\zoomDepth}[4]{
        \trimbox{0cm 0cm 0.15cm 0.0cm}{
        \begin{tikzpicture}[
    		image/.style={inner sep=0pt, outer sep=0pt},
    		collabel/.style={above=9pt, anchor=north, inner ysep=0pt, align=center}, 
    		rowlabel/.style={left=9pt, rotate=90, anchor=north, inner ysep=0pt, scale=0.8, align=center},
    		subcaption/.style={inner xsep=0.75mm, inner ysep=0.75mm, below right},
    		arrow/.style={-{Latex[length=2.5mm,width=4mm]}, line width=2mm},
    		spy using outlines={rectangle, size=1.25cm, magnification=2.5, connect spies, ultra thick, every spy on node/.append style={thick}},
    		style1/.style={cyan!90!black,thick},
    		style2/.style={orange!90!black},
    		style3/.style={blue!90!black},
    		style4/.style={green!90!black},
    		style5/.style={white},
    		style6/.style={black},
        ]
        
        \node [image] (#1) {\imgDepth[0px 0px 0px 0px]{#2}};
        \spy[style1] on ($(#1.center)-#3$) in node (crop-#1) [anchor=#4] at (#1.#4);
        
        \end{tikzpicture}
        }
    }

    \vspace{-2mm}
    \setlength{\tabcolsep}{0.5pt}
    \begin{tabular}{cccccc}
    &
    Input &
    Ours &
    TöRF \cite{attal2021torf} &
    DeformableGS \cite{yang2024deformable} &
    DeformableGS (--depth) 
    \\
    
    \multirow{2}*{\raisebox{0.0in}{\rotatebox[origin=c]{90}{\SeqCupboard}}} &
    \zoomDepth{label1}{images/qualitative_comparison/torf_real/cupboard/color_0_input_0020.png}{(\zoomxCupboard+0.2,\zoomyCupboard+0.1)}{north west} &
    \zoomDepth{label2}{images/qualitative_comparison/torf_real/cupboard/color_1_ours_0020.png}{(\zoomxCupboard,\zoomyCupboard)}{north west} &
    \zoomDepth{label3}{images/qualitative_comparison/torf_real/cupboard/color_2_torf_0020.png}{(\zoomxCupboard,\zoomyCupboard)}{north west} &
    \zoomDepth{label4}{images/qualitative_comparison/torf_real/cupboard/color_3_rgbd_0020.png}{(\zoomxCupboard,\zoomyCupboard)}{north west} &
    \zoomDepth{label5}{images/qualitative_comparison/torf_real/cupboard/color_4_rgb_0020.png}{(\zoomxCupboard,\zoomyCupboard)}{north west} \\
    &
    \zoomDepth{label1}{images/qualitative_comparison/torf_real/cupboard/depth_0_input_0020.png}{(\zoomxCupboard,\zoomyCupboard)}{north west} &
    \zoomDepth{label2}{images/qualitative_comparison/torf_real/cupboard/depth_1_ours_0020.png}{(\zoomxCupboard,\zoomyCupboard)}{north west} &
    \zoomDepth{label3}{images/qualitative_comparison/torf_real/cupboard/depth_2_torf_0020.png}{(\zoomxCupboard,\zoomyCupboard)}{north west} &
    \zoomDepth{label4}{images/qualitative_comparison/torf_real/cupboard/depth_3_rgbd_0020.png}{(\zoomxCupboard,\zoomyCupboard)}{north west} &
    \zoomDepth{label5}{images/qualitative_comparison/torf_real/cupboard/depth_4_rgb_0020.png}{(\zoomxCupboard,\zoomyCupboard)}{north west} \\

    \multirow{2}{*}{\raisebox{0.0\height}{\rotatebox[origin=c]{90}{Deskbox}}} &
    \zoomDepth{label1}{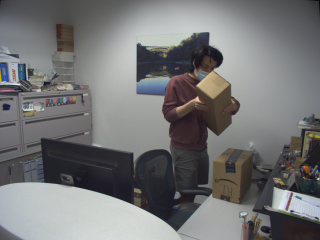}{(\zoomxDeskbox-0.1,\zoomyDeskbox+0.1)}{north west} &
    \zoomDepth{label2}{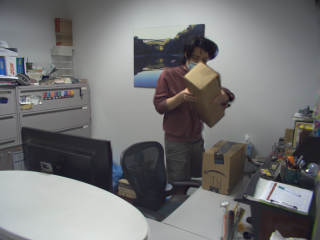}{(\zoomxDeskbox,\zoomyDeskbox)}{north west} &
    \zoomDepth{label3}{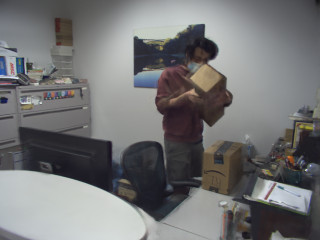}{(\zoomxDeskbox,\zoomyDeskbox)}{north west} &
    \zoomDepth{label4}{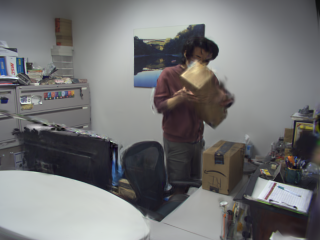}{(\zoomxDeskbox,\zoomyDeskbox)}{north west} &
    \zoomDepth{label5}{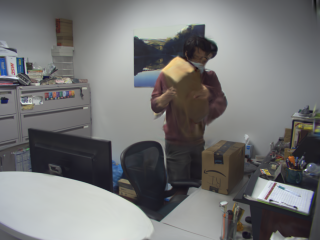}{(\zoomxDeskbox,\zoomyDeskbox)}{north west} \\
    [-1.5em] &
    \zoomDepth{label1}{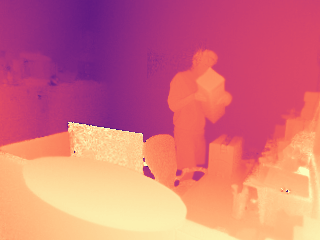}{(\zoomxDeskbox-0.1,\zoomyDeskbox)}{north west} &
    \zoomDepth{label2}{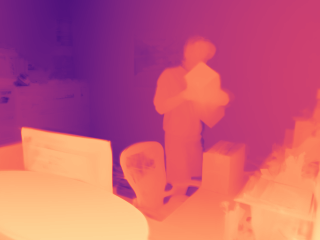}{(\zoomxDeskbox,\zoomyDeskbox)}{north west} &
    \zoomDepth{label3}{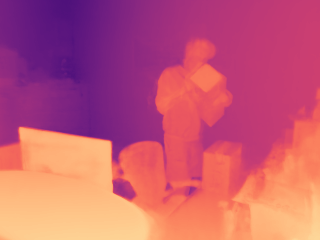}{(\zoomxDeskbox,\zoomyDeskbox)}{north west} &
    \zoomDepth{label4}{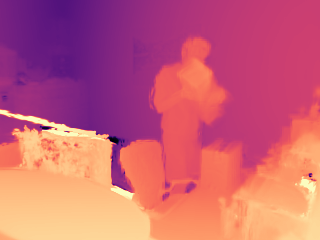}{(\zoomxDeskbox,\zoomyDeskbox)}{north west} &
    \zoomDepth{label5}{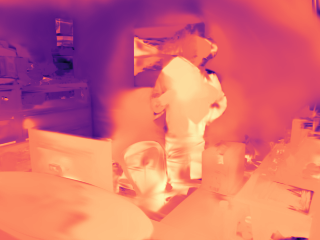}{(\zoomxDeskbox,\zoomyDeskbox)}{north west} \\
    
    \multirow{2}*{\raisebox{0.0in}{\rotatebox[origin=c]{90}{\SeqPhotocopier}}} &
    \zoomDepth{label1}{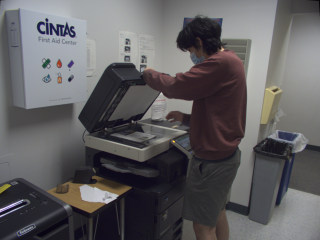}{(\zoomxCopier+0.3,\zoomyCopier)}{north west} &
    \zoomDepth{label2}{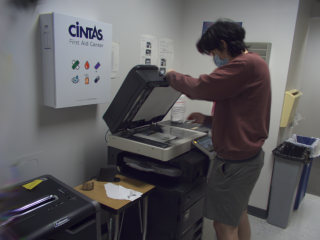}{(\zoomxCopier,\zoomyCopier)}{north west} &
    \zoomDepth{label3}{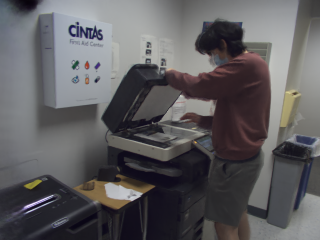}{(\zoomxCopier,\zoomyCopier)}{north west} &
    \zoomDepth{label4}{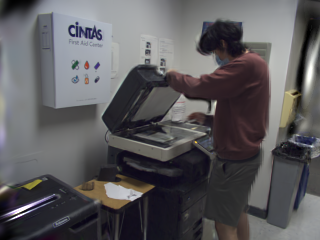}{(\zoomxCopier,\zoomyCopier)}{north west} &
    \zoomDepth{label5}{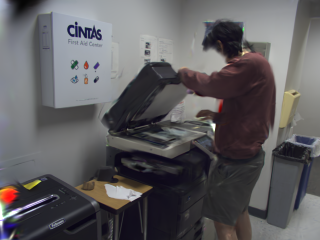}{(\zoomxCopier,\zoomyCopier)}{north west} \\
    &
    \zoomDepth{label1}{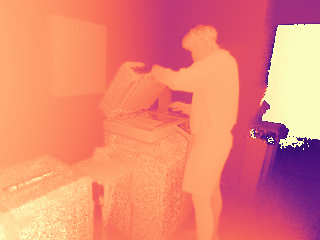}{(\zoomxCopier+0.25,\zoomyCopier)}{north west} &
    \zoomDepth{label2}{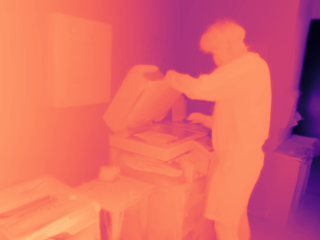}{(\zoomxCopier,\zoomyCopier)}{north west} &
    \zoomDepth{label3}{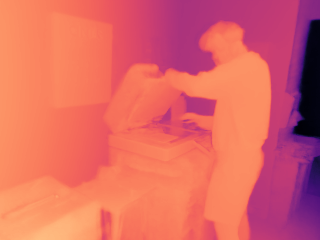}{(\zoomxCopier,\zoomyCopier)}{north west} &
    \zoomDepth{label4}{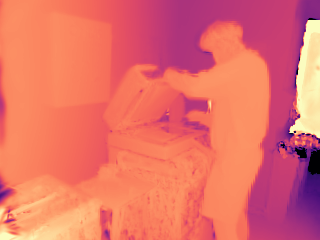}{(\zoomxCopier,\zoomyCopier)}{north west} &
    \zoomDepth{label5}{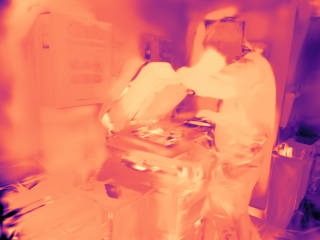}{(\zoomxCopier,\zoomyCopier)}{north west} \\

    \multirow{2}{*}{\raisebox{0.0\height}{\rotatebox[origin=c]{90}{\SeqStudyBook}}} &
    \zoomDepth{label1}{images/qualitative_comparison/torf_real/studybook/color_0_input_0002.png}{(\zoomxStudybook+0.18,\zoomyStudybook-0.05)}{north east} &
    \zoomDepth{label2}{images/qualitative_comparison/torf_real/studybook/color_1_ours_0002.png}{(\zoomxStudybook,\zoomyStudybook)}{north east} &
    \zoomDepth{label3}{images/qualitative_comparison/torf_real/studybook/color_2_torf_0002.png}{(\zoomxStudybook,\zoomyStudybook)}{north east} &
    \zoomDepth{label4}{images/qualitative_comparison/torf_real/studybook/color_3_rgbd_0002.png}{(\zoomxStudybook,\zoomyStudybook)}{north east} &
    \zoomDepth{label5}{images/qualitative_comparison/torf_real/studybook/color_4_rgb_0002.png}{(\zoomxStudybook,\zoomyStudybook)}{north east} \\
    &
    \zoomDepth{label1}{images/qualitative_comparison/torf_real/studybook/depth_0_input_0002.png}{(\zoomxStudybook+0.13,\zoomyStudybook-0.18)}{north east} &
    \zoomDepth{label2}{images/qualitative_comparison/torf_real/studybook/depth_1_ours_0002.png}{(\zoomxStudybook,\zoomyStudybook)}{north east} &
    \zoomDepth{label3}{images/qualitative_comparison/torf_real/studybook/depth_2_torf_0002.png}{(\zoomxStudybook,\zoomyStudybook)}{north east} &
    \zoomDepth{label4}{images/qualitative_comparison/torf_real/studybook/depth_3_rgbd_0002.png}{(\zoomxStudybook,\zoomyStudybook)}{north east} &
    \zoomDepth{label5}{images/qualitative_comparison/torf_real/studybook/depth_4_rgb_0002.png}{(\zoomxStudybook,\zoomyStudybook)}{north east} \\

    \end{tabular}
    \vspace{-2mm}
    \caption{\textbf{More results on \torf real-world scenes}. DeformableGS is given C-ToF-derived depth as an additional input.}
    \label{fig:torf-qualitative-baselines-supp}
\end{figure*}
\begin{figure*}[htpb]
    \centering
    \setlength{\tabcolsep}{0.5cm}
    \begin{tabular}{ccc}
        \includegraphics[width=0.3\textwidth]{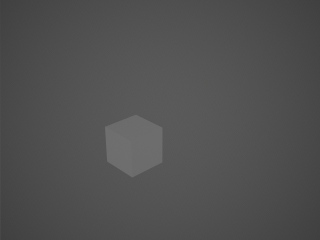} &
        \includegraphics[width=0.3\textwidth]{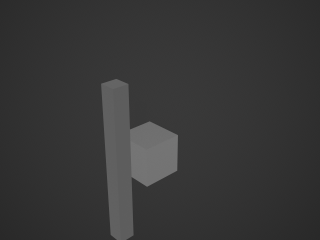} &
        \includegraphics[width=0.3\textwidth]{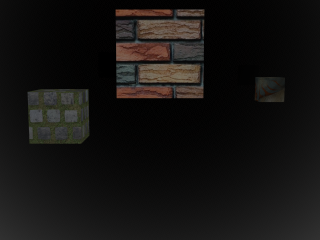} \\
        (a) \SeqSlidingCube & (b) \SeqOccludedCube & (c) \SeqZMotionShort \\[0.5cm]
        
        \includegraphics[width=0.3\textwidth]{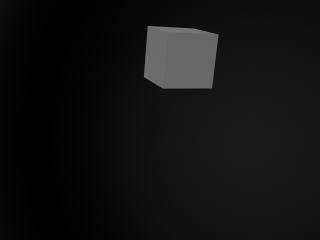} &
        \includegraphics[width=0.3\textwidth]{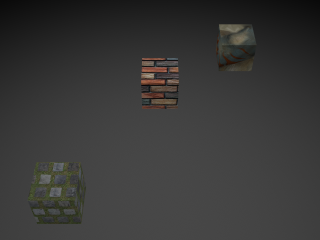} &
        \includegraphics[width=0.3\textwidth]{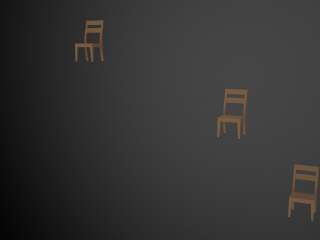} \\
        (d) \SeqArcingCube & (e) \SeqAxialShort & (f) \SeqThreeChairsShort \\[0.5cm]
        
        \includegraphics[width=0.3\textwidth]{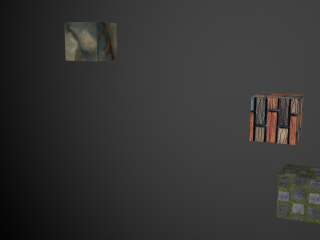} &
        & \\
        (g) \SeqThreeCubesShort & & \\
    \end{tabular}

    \caption{Corresponding RGB color images for the synthetic scenes for \mainfig{4}
    and \cref{fig:real-synthetic-baselines-supp}. Note that \SeqSlidingCube, \SeqOccludedCube, and \SeqArcingCube have no texture.}
    \label{fig:synthetic_rgb_legend}
\end{figure*}

\end{document}